\documentclass[graybox]{svmult}

\usepackage{helvet}
\usepackage{courier}
\usepackage{type1cm}
\usepackage{makeidx}
\usepackage{graphicx}

\usepackage{multicol}
\usepackage[bottom]{footmisc}
\usepackage{hyperref}
\usepackage{soul}
\usepackage{amsmath}
\usepackage{siunitx}
\usepackage{upgreek}
\hypersetup{colorlinks=true,urlcolor=blue}
\usepackage[square,numbers]{natbib}

\makeindex

\begin{document}
\title*{Quantum sensors with matter waves for GW observation}
\author{Andrea Bertoldi, Philippe Bouyer and Benjamin Canuel}
\institute{Andrea Bertoldi \at Laboratoire Photonique, Num{\'e}rique et Nanosciences (LP2N), Universit{\'e} Bordeaux - IOGS - CNRS:UMR 5298, 1 rue Fran{\c c}ois Mitterrand, 33400 Talence, France, \email{andrea.bertoldi@institutoptique.fr}
\and Philippe Bouyer \at Laboratoire Photonique, Num{\'e}rique et Nanosciences (LP2N), Universit{\'e} Bordeaux - IOGS - CNRS:UMR 5298, 1 rue Fran{\c c}ois Mitterrand, 33400 Talence, France, \email{philippe.bouyer@institutoptique.fr}
\and Benjamin Canuel \at Laboratoire Photonique, Num{\'e}rique et Nanosciences (LP2N), Universit{\'e} Bordeaux - IOGS - CNRS:UMR 5298, 1 rue Fran{\c c}ois Mitterrand, 33400 Talence, France, \email{benjamin.canuel@institutoptique.fr}}

\maketitle

\abstract{Quantum sensors exploiting matter waves interferometry promise the realization of a new generation of Gravitational Wave detectors. The intrinsic stability of specific atomic energy levels makes atom interferometers and clocks ideal candidates to extend the frequency window for the observation of Gravitational Waves in the mid-frequency band, ranging from 10 mHz to 10 Hz. We present the geometry and functioning of this new class of ground and space detectors and detail their main noise sources. We describe the different projects undertaken worldwide to realize large scale demonstrators and push further the current limitations. We finally give the roadmap for achieving the instrumental sensitivity required to seize the scientific opportunities offered by this new research domain.}

\section{Keywords} 
Atom Interferometry, Cold Atoms, Gravity Gradiometry, Gravity-Gradient Noise, Mid-band Gravitational Wave Detection, Atom-Laser Antenna, Multiband GW Astronomy, Cold Atoms, Quantum Sensors

\section{Introduction}

The observation of Gravitational Waves (GWs)~\cite{Abbott2020} has opened a new era of GW astronomy that can bring new insight for the study of general relativity in its most extreme regimes, dark matter or the exploration of the early universe, where light propagation was impossible. It is becoming possible to study a large range of GW sources and frequencies, from well understood phenomena~\cite{Mandel2018}, to more speculative ones~\cite{Nielsen1973,Weir2018}.

To widen the reach of GW astronomy it is necessary to explore a frequency range beyond that accessible by state-of-the-art detectors \cite{Aasi2015,Acernese2014}, which currently spans from 10 Hz to 10 kHz~\cite{Sesana2016}. While planned detectors will either push GW sensitivity in the current band with the third generation ground-based laser interferometer (Einstein Telescope - ET) ~\cite{Punturo2010,Abbott2017}, or investigate GWs sources at very low frequency with the space-based Laser Interferometer Space Antenna (LISA) ~\cite{Jennrich2009}, the critical infrasound (0.1 Hz-10 Hz) band \cite{Mandel2018,Kuns2020,Sedda2020,ElNeaj2020} is left open to new concepts. Sensors based on quantum technologies, such as atom interferometers (AIs) \cite{Dimopoulos2008b,Canuel2020} or atomic clocks \cite{Kolkowitz2016,Rudolph2020}, can provide a plausible answer to this challenge.

The continuous development over more than three decades of techniques to manipulate and coherently control ultracold atomic samples  has pushed theses sensors to unprecedented levels of accuracy and stability. Atomic clocks reach today a stability at about the 10$^{-18}$ level \cite{Hong2005,Katori2003,Jiang2011,Ludlow2015,Beloy2021,Diddams2020,Mackenzie2020,Takamoto2020}, and their precision continues to improve so that one day it will be feasible, for instance, to utilize them for a direct detection of the gravitational field \cite{Hollberg2017}. Atom optics and matter waves manipulation also pushed the development of new generation of force sensors exhibiting unprecedented sensitivity and accuracy \cite{Barrett2014,Geiger2020}. They can nowadays address many applications, such as probing inertial forces \cite{Peters1999,Snadden1998,Gustavson1997}, studying fundamental physics and testing gravitational theories \cite{Tino2014,Bongs2019,Lorek2012,Tino2019,Schlippert2020,Burrage2015}.

The availability of these sensors motivated the emergence of concepts for ultra high precision measurements of time and space fluctuations, with direct application to tests of general relativity \cite{Antoine_2003,Wolf2011,Dimopoulos2008b} and in particular with a potential to open an inaccessible window in GW detection. While the first proposals \cite{Chiao2003} were purely speculative, concepts have recently evolved to extended proposals for space \cite{Graham2013,Kolkowitz2016} or ground \cite{Canuel2018,Coleman2019} with the current developments of lower scale version of what could be a future generation infrasound GW detector.

\section{Atom interferometry and GW detection}

\subsection{Principle of GW detection using matter waves interferometry}Since the first AIs were realized almost three decades ago, these elegant experimental demonstrations of quantum physics have evolved to instruments at the leading edge of precision measurements. They allow for measuring inertial or gravitational forces affecting the propagation of matter waves with a sensitivity and precision comparable to or even better than existing classical sensors. Their present performance and technological maturity provide breakthrough capacities in a variety of fields \cite{Bongs2019} from applied to fundamental sciences \cite{Barrett2014} such as navigation and gravimetry. They are nowadays developed both by academic teams and industry, with specific focus in  miniaturization \cite{Chen2019}, transportability \cite{Barrett2016,Menoret2018} without compromising their performances. Examples are gyroscopes, gravimeters, gradiometers, with applications in navigation, geophysics, metrological determination of fundamental constants, and tests of GR. The reported precision and sensitivity of these inertial sensors are parts in $10^9$ of Earth gravity and rotation rates at the verge of $10^{-9}$ rad/s/$\sqrt{\rm Hz}$. These values compare favorably with current technologies, even outside the very quiet environment of a laboratory. 

\begin{figure}[hbt!]
\centering
\includegraphics[width=.75\linewidth]{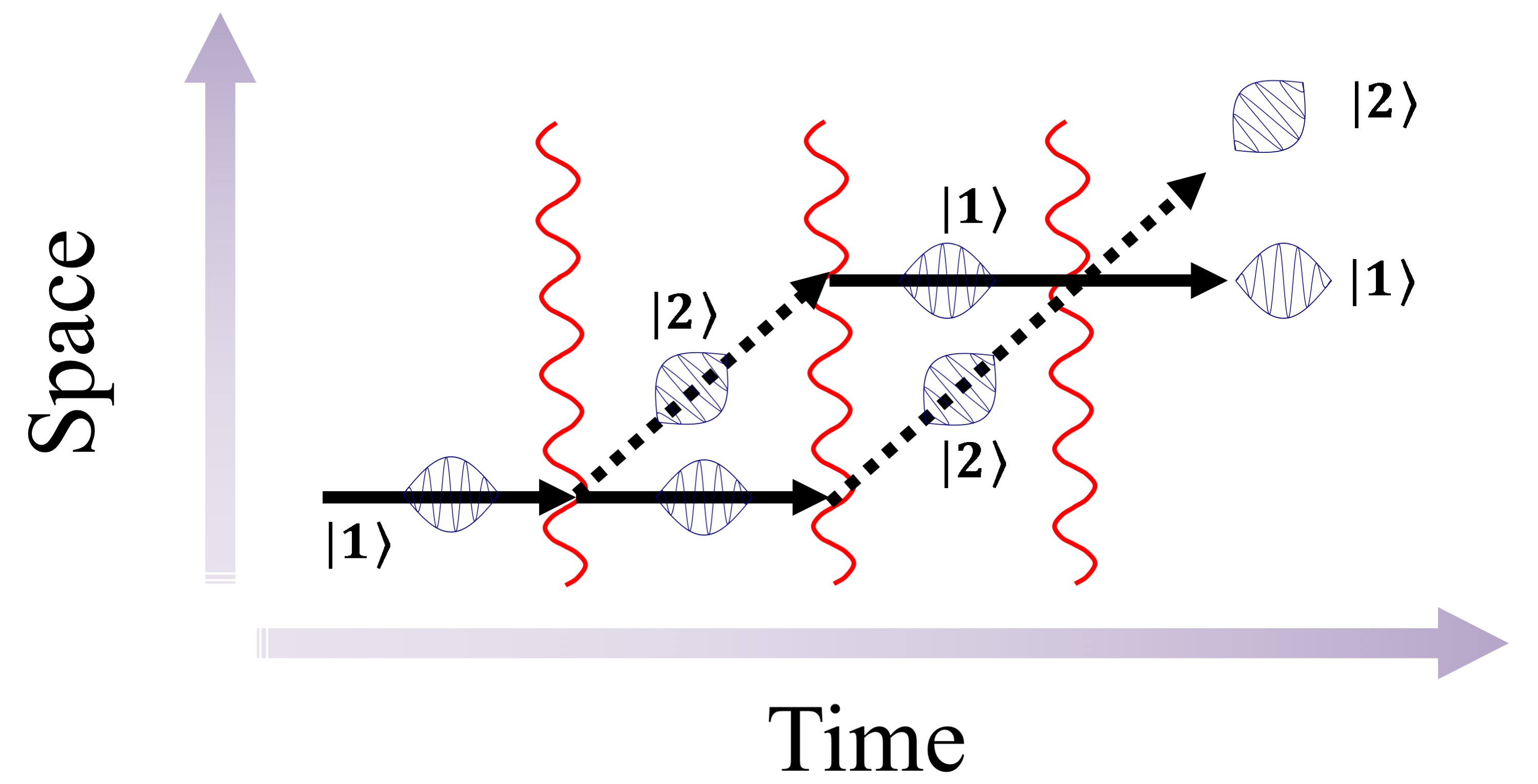}
\caption{Space-time diagram schematic of an AI using light pulses. The atomic trajectories are represented in black: the solid lines refer to the propagation in state $\left|1\right>$, the dashed ones in state $\left|2\right>$. The propagation paths are represented as straight lines, whereas uniform gravity makes them parabolic. The two states have a momentum separation given by the two-photon momentum exchange imparted by the interferometric pulses, represented by the sinusoidal red lines. Taken from \cite{Canuel2020} under CC BY 4.0.
\label{fig:spacetime}} 
\end{figure}

In general, an AI uses a succession of light pulses tuned to a particular atomic frequency resonance, that act as coherent beam splitters separated by a time $T$ \cite{Kasevich1991}. The first pulse splits the incoming matter wave into two wavepackets that follow different paths. In direct analogy with light, the accumulation of phase along these two paths leads to interference at the last beam-splitter. Each output channels will then exhibit complementary probability as a sinusoidal function of the accumulated phase difference, $\Delta \phi$. Most AIs follow the Mach-Zehnder design: two beam splitters with a mirror inserted inside to fold the paths (see Fig. \ref{fig:spacetime}). The sensitivity of such an interferometer is defined by the area enclosed by the two atomic trajectories. When atoms are subject to acceleration or rotation along their trajectory, their speed along this trajectory is modified, which modifies their de Broglie wavelength and ultimately leads to a variation of $\Delta \phi$. The output of the AI blends then together the effects of rotation and acceleration, as well as unwanted contributions from wave-front distortions and mirror vibrations. For instance, if the platform containing the laser beams accelerates, or if the atoms are subject to an acceleration $a$, the phase shift becomes:
\begin{equation}
  \label{eq:accphase}
  \Delta\phi_{\rm acc} = {\rm k}_{\rm eff} a  T^2,
\end{equation}
where ${\rm k}_{\rm eff}$ is the effective wave number of the coherent manipulation laser. When the laser beams are vertically directed, the interferometer measures the acceleration due to gravity $g$. Remarkably, ppb-level sensitivity, stability and accuracy has been achieved with such gravimeter \cite{Geiger2020}.

Combining two such interferometers separated in space and using a common laser beam is well-suited to measure gravity gradient\cite{Snadden1998} as a results of the differential phase shift. In this way, major technical background noises are common-mode rejected, which leads to nearly identical phase shifts if each interferometer is subjected to the same acceleration. This configuration can measure Earth's gravity gradient, as well as the modification of gravity from nearby mass distributions \cite{Bertoldi2006}. In the laboratory, gravity gradiometers have achieved resolutions below $10^{-9}$ s$^{-2}$, and allowed the precise determination of the gravitational constant $G$ \cite{Fixler2007,Lamporesi2008,Rosi2014}. 

\begin{figure}[htp]
\centering
\includegraphics[width=.75\linewidth]{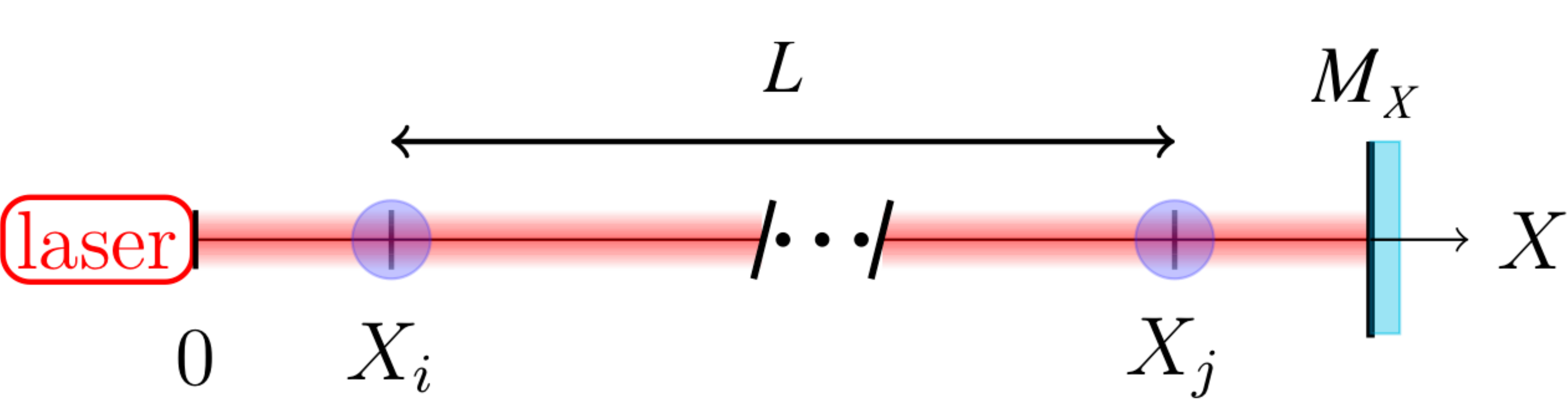}
\caption{Atom interferometry based GW detector using a gradiometric configuration of baseline $L$. The AIs are placed at $X_{i,j}$. A common interrogation laser is retro-reflected by a mirror at position $M_X$. Taken from \cite{Canuel2020} under CC BY 4.0. \label{fig:AIgradio}}
\end{figure}

The dependence of $\Delta \phi$ to inertia can more generally be seen as a direct consequence of the propagation of matter waves in curved space-time \cite{Antoine_2003,Asenbaum2017}, thus leading to consider atom interferometry as a potential candidate for precision tests of general relativity and consequently for GW detection \cite{Chiao2003}. A GW affecting the AI would typically induce a  phase shift $\Delta \phi\sim h(t) T{\rm v}{\rm k}_{\rm eff}/2\pi$, where $h(t)$ is the GW strain amplitude and v the velocity of the atoms entering the interferometer \cite{Delva2007}. For this effect to be large considering a single atom, the instrument would have to be of unreasonable size (the largest AIs are of meter scales, whereas hundreds of meters would be needed) and use high velocity (besides the fact that sensitivity improves with cold-slow atoms). These limiting factors vanish when the AI is not solely used to read out the GW dephasing, but to differentially measure the effect of the GW on the propagation of light (see Fig. \ref{fig:AIgradio}), either by measuring how the light propagation time can be affected \cite{Kolkowitz2016} or how the laser phase can be modulated \cite{Dimopoulos2008b}. This connects directly to the current GW detectors based on laser interferometry \cite{Acernese2014,Aasi2015,Jennrich2009}, but with the use of matter-waves acting as quasi-perfect, free-falling, proof masses. The resulting spacecraft requirements for space-based missions are significantly reduced, and access is granted, on Earth, to frequencies usually hindered by seismic and more general geophysical noise \cite{Canuel2018,Har2015, Chaibi2016}.

\subsection{Space-based and terrestrial instruments}

There is a strong interest, and expected synergies to ultimately use concomitantly both terrestrial and space-based GW detectors. Today, GW observation is performed thanks to three running ground based detectors and new ones are currently under development or nearly ready. Location on the ground provides a ideal environment to develop and further enhance novel concepts to either improve current instruments (such as the use of quantum states of light in LIGO \cite{Aasi2013} or VIRGO) or develop new ones \cite{Punturo2010,Akutsu2018}. There are two major challenges in order to improve the detectors sensitivity and expand it to lower frequency.

\begin{figure}[htp]
\centering
\includegraphics[height=150pt]{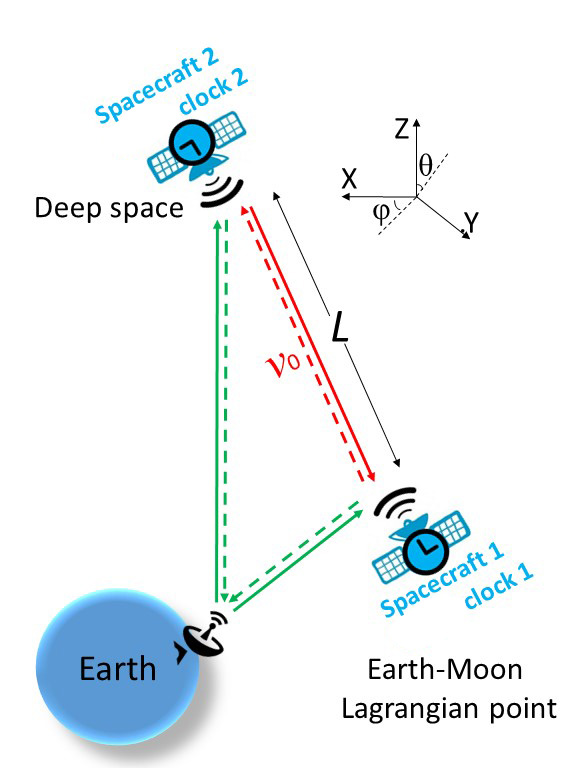}
\hspace{.75 in}
\includegraphics[height=150pt]{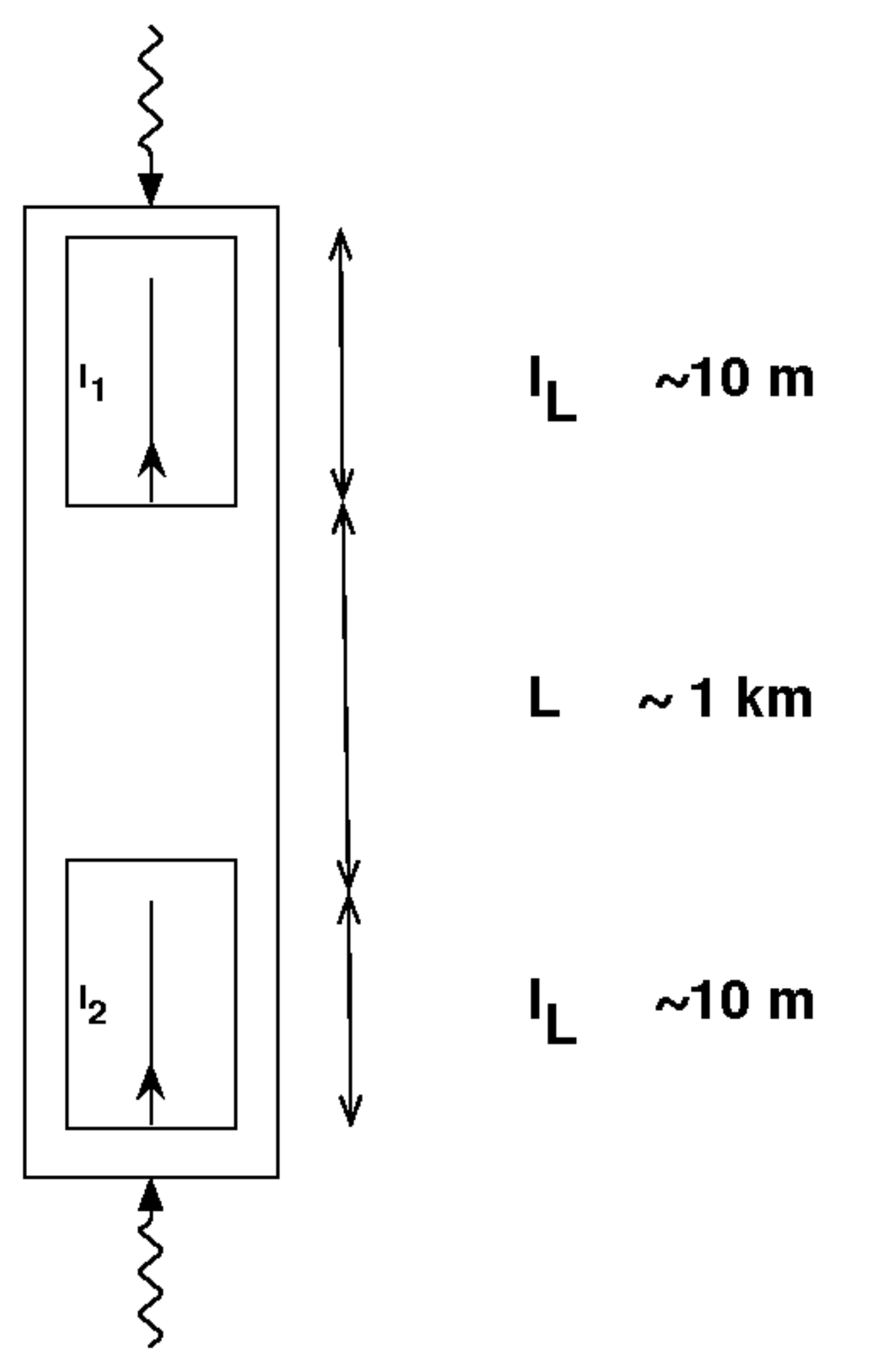}
\caption{Space and ground based detectors. (left) In the DOCS proposal \cite{Su2018} spacecraft 1 and 2 are set to one of the Earth-Moon Lagrangian points and to deep space ($L$ = 1.5 AU), respectively. Each of the spacecrafts has an optical clock on board. A radio signal with a nominal frequency $\nu_0$ is transmitted from spacecraft 1 (or 2) to spacecraft 2 (or 1), namely using two one-way link. Taken from \cite{Su2018} under CC BY 3.0. (right) Diagram of the setup for a terrestrial experiment, taken with permission from \cite{Dimopoulos2008}, copyright by the American Physical Society. The straight lines represent the path of the atoms in the two $I_L \sim 10$~m  interferometers $I_{1}$ and $I_{2}$ separated vertically by $L \sim 1$~km. The wavy lines represent the paths of the lasers.
\label{fig:spaceground}}
\end{figure}

First, vibrations (or seismic noise) needs to be mitigated. In LIGO/VIRGO, this is achieved by levitating the mirror proof masses thanks to suspension systems which set the current lower band limit to about 10 Hz. Atom interferometry solves this problem by using free falling atoms (that are naturally isolated from vibrations) and by tuning the AI so that its maximum sensitivity is at the desired low frequency~\cite{Dimopoulos2008b,Graham2016}. Second, even without vibrations (for instance by setting the instrument deep underground), the fluctuations of mass around the instrument lead to fluctuations of the gravitational forces that typically forbid any signal to be detected below a few Hz \cite{Har2015}. This \textit{Gravity-Gradient Noise} (GGN) or \textit{Newtonian Noise} (NN) may finally be mitigated by mapping it using a network of ultra precise accelerations sensors that can correlate the noise to the measurement itself~\cite{Chaibi2016}.

Space, on the other hand, provides the perfect environment to detect and monitor GW at very low frequencies. Spacecraft can be set \textit{drag-free} to be in perfect free fall and GW signals can be extracted by monitoring the propagation time of the light between two satellites over very large distance (kilometers to millions of kilometers), as shown in Fig. \ref{fig:spaceground}. This gives access to very low frequency (mHz to Hz) GW signatures that can be precursors of larger signals later monitored by ground based detectors \cite{Sedda2020}. Space provides also an ideal environment for atom interferometry, since the lack of gravity opens for enhanced sensitivities, with larger interrogation times $T$ and interferometer arms separations ${\rm k}_{\rm eff} T$ that approach the initial ideas of \cite{Chiao2003}.  

\subsection{Classes of quantum-sensor based gravitational wave detector \label{sec:sensor2photons}}

Before performing a phase measurement with an AI, many steps must be performed \cite{Bongs2019}: prepare an ensemble of cold atoms with the proper velocity distribution, i.e. temperature, and state, perform the appropriate sequence of matter-wave interferometry and finally read-out the interferometer signal and extract the information about the phase.

A light-pulse interferometer sequence uses, to produce the interference, a series of laser pulses applied on atoms that follow free-fall trajectories (see Fig.~\ref{fig:spacetime}). This process relies on the exchange of momentum $\hbar{\rm k}_{\rm eff}$ between the atoms and the lasers, while at the same time avoiding to drive spontaneous emission from the laser excitation. Several schemes can achieve this. They either rely on two-photon processes - where the atoms are always in their ground, non-excited states - such as Bragg diffraction \cite{Giltner1995} or two-color Raman processes \cite{Kasevich1991} or on single-photon excitation of a nearly forbidden, long-lived transition \cite{Graham2013}.

\subsubsection{Two-photon transition based interferometers}

\begin{figure}
\begin{center}
\includegraphics[height=100pt]{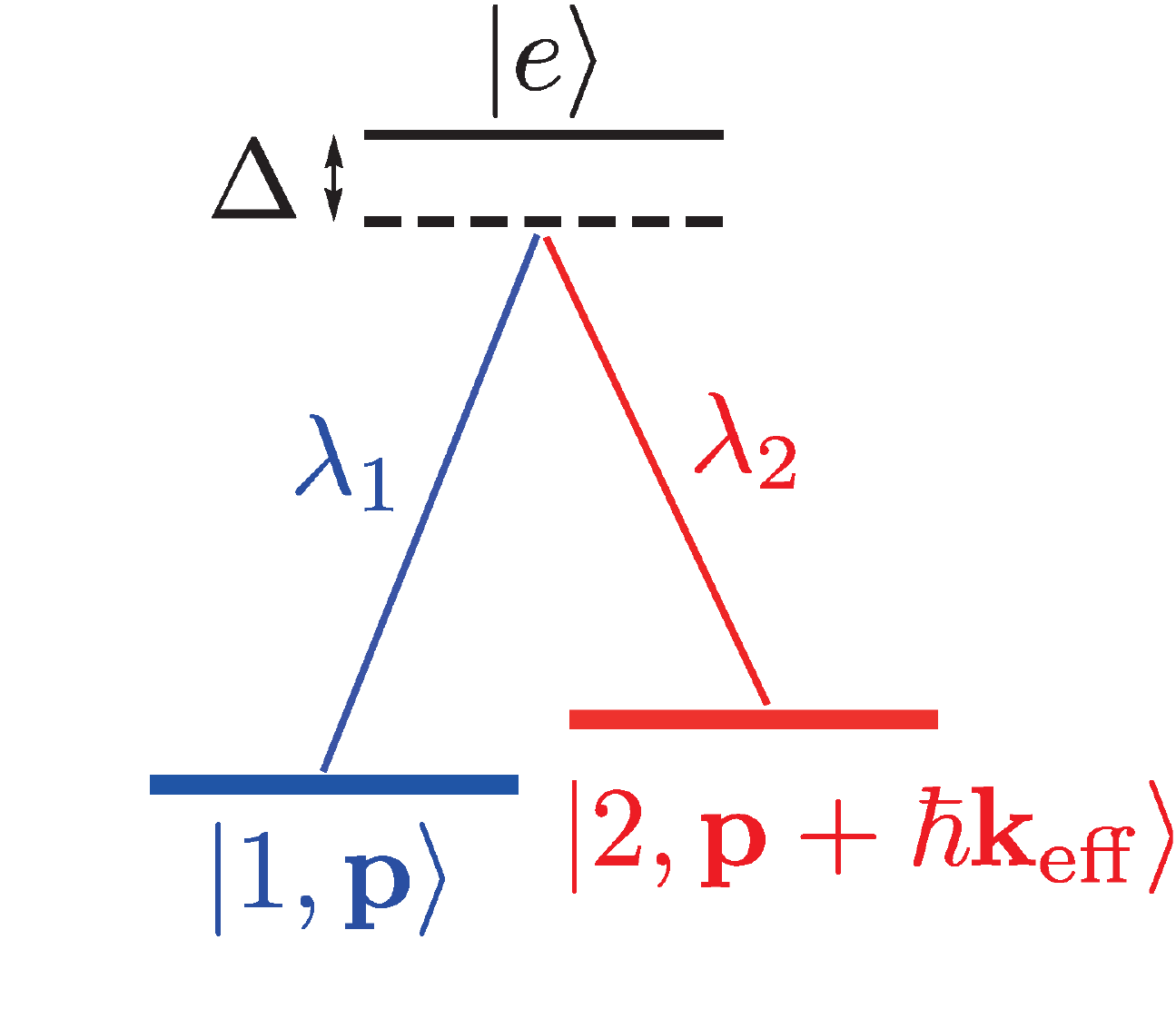}
\hspace{.2 in}
\includegraphics[height=100pt]{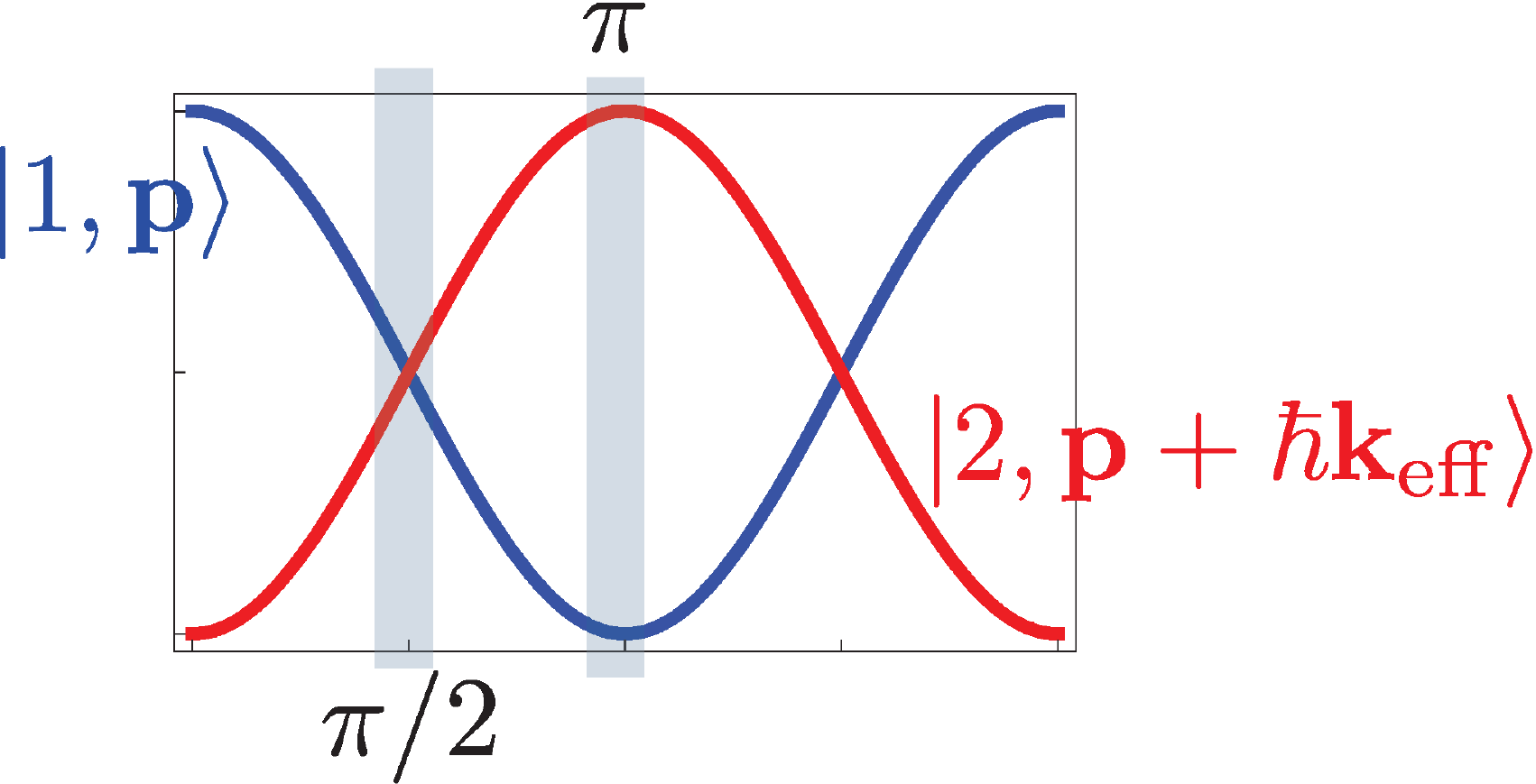}
\caption{(left) Stimulated Raman transition between atomic states $\left|1\right>$ and $\left|2\right>$ using lasers of wavelength $\lambda_1$ and $\lambda_2$. (right) Rabi oscillations between states $\left|1\right>$ and $\left|2\right>$.  A $\frac{\pi}{2}$ pulse is a beamsplitter since the atom ends up in a superposition of states $\left|1\right>$ and $\left|2\right>$ while a $\pi$ pulse is a mirror since the atom changes from state $\left|1\right>$ to state $\left|2\right>$.}
\label{Fig:Raman}
\end{center}
\end{figure}

Bragg and Raman two-photon diffraction processes both rely on the same principle: two lasers are far detuned from an optical transition, and their frequency difference is set equal to the atom's recoil kinetic energy plus any internal energy shift. During a pulse of light, the atom undergoes a Rabi oscillation between the two states $\left|1,\mathbf{p}\right>$ and $\left|2,\mathbf{p}+\mathbf{k}_{\rm eff}\right>$). A beamsplitter results when the laser pulse time is equal to a quarter of a Rabi period ($\frac{\pi}{2}$ pulse), and a mirror requires half a Rabi period ($\pi$ pulse). In the specific case of Raman diffraction, the internal and external degrees of freedom of the atom are entangled, resulting in an energy level change and a momentum kick (see Fig. \ref{Fig:Raman})

The typical AI configuration of Fig. \ref{fig:spacetime} follows a beamsplitter-mirror-beamsplitter $(\frac{\pi}{2}-\pi-\frac{\pi}{2})$ sequence \cite{Kasevich1991}: the initial beamsplitter ($\frac{\pi}{2}$) pulse creates a superposition of states which differ in velocity by $\mathbf{k}_{\rm eff}/m$. The resulting spatial separation after a time of flight $T$ sets the interferometer's sensitivity to gravity along the direction of $\mathbf{k}_{\rm eff}$.  The mirror ($\pi$) pulse reverses the relative velocity and the final ($\frac{\pi}{2}$) pulse, applied at time $2 T$, interferes these overlapping components. The interferometric fringes are then detected using light induced fluorescence detection. 

If the detection is limited only by the quantum projection noise of the atoms (atom shot noise), the phase differences $\Delta\phi$ can be measured with a sensitivity below $10^{-3}~\text{rad}/\sqrt{\text{Hz}}$, which corresponds to the shot noise limit $1/\sqrt{N}$ for $10^6$ atoms in the interferometer. Higher atom number, or the use of squeezed atomic states instead of uncorrelated thermal atom ensembles can increase further this sensitivity. If two light-pulse interferometer sequences are performed simultaneously at different positions (see Fig.~\ref{fig:AIgradio}) in a gradiometer configuration, the gravitational wave of strain amplitude $h$ and frequency $\omega$ will typically produce a differential acceleration signal $\sim hL\omega^2$ on the interferometers, as shown in detail in Sec.~\ref{sec:noise}. A phase sensitivity of $10^{-5}~\text{rad}/\sqrt{\text{Hz}}$ can target an acceleration sensitivity of $10^{-15}~g/\sqrt{\text{Hz}}$ leading to a strain sensitivity of $10^{-18}/\sqrt{\text{Hz}}$ for $L=1$~km.

Each interferometer sequence must be repeated in order to record the GW signal. The measurement repetition rate will put limits of the GW detection frequency range; usually it is limited by the time to produce the sample of cold atoms, and can thus increase when using colder samples. On earth, typical repetition rates of $10~\text{Hz}$ or higher must be achieved for allowing to target frequencies up to $5 \text{~Hz}$ without reaching sampling problems.  In space, where target frequencies can be as low as  $\sim 10^{-3} - 1 \text{~Hz}$, the sampling rate can be reduced below $1~\text{Hz}$. To fulfill this requirement interleaved measurement sequences will be adopted \cite{Biedermann2013,Savoie2018}.

\begin{figure}
    \centering
    \includegraphics[width=0.75\linewidth]{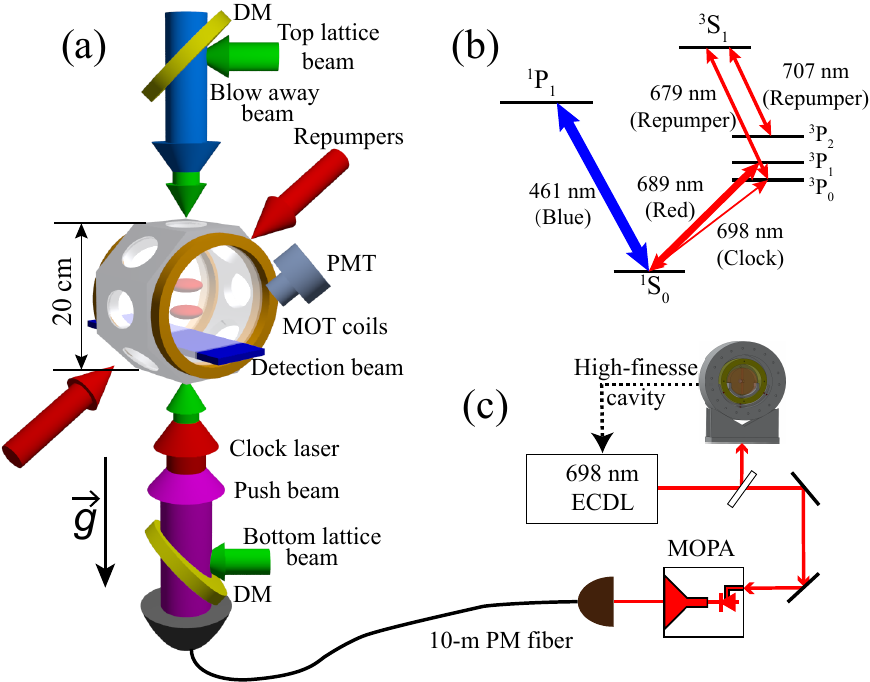}
    \caption{Experimental apparatus of an ultra-cold strontium AI (a) using interferometry pulses on the clock transition at 698 nm shown in red on the energy diagram (b). Laser radiation at 698 nm is frequency stabilized in two steps to ultra-stable optical cavities (c). MOT: magneto-optical trap; PDH: Pound-Drever-Hall; MOPA: master oscillator power amplifier; PM: polarization-maintaining; ECDL: external-cavity diode laser; DM: dichroic mirror; PMT: photomultiplier tube. Taken from \cite{Hu2020} under CC BY 3.0.}
     \label{fig:srAIflorence}
\end{figure}

As for light-based interferometer detection, laser frequency noise is one limiting factor in this gradiometer configuration. Usually, a configuration of two orthogonal interferometer arms can exploit the quadrupolar nature of gravitational radiation to separate gravitational wave induced phase shifts from those arising from laser noise. For the gradiometer configuration of Fig.~\ref{fig:AIgradio}), laser frequency noise is suppressed since the same laser beams interrogate both ensembles of atoms along a common line-of-sight.  Nevertheless, the time delay between the two AIs and the need of two counter-propagating laser beams for each AIs \cite{Snadden1998} leaves a residual sensitivity to laser frequency noise, or to optical elements vibrations, and the impact of this effect increases with the baselines length $L$. One solution consists in eliminating the retro-reflected laser beams altogether by driving optical atomic transitions with a single laser \cite{Graham2013}, as recently demonstrated in \cite{Hu2017} (Fig. \ref{fig:srAIflorence}).

\subsubsection{Optical clocks and single photon based interferometers}

In contrast to Raman or Bragg two-photon diffraction, driving a narrow-linewidth clock transitions only requires a single resonant laser beam. In these clock transitions, for instance the strontium ${}^1S_0~\rightarrow {}^3P_0$ one, the spontaneous emission loss from excited state decay can be neglected thanks to the $150~\text{s}$ lifetime~\cite{Santra2004}, and if needed minimized by evolving both paths of the interferometer in the ground state by means of extra pulses \cite{Graham2013}. As for optical atomic clocks \cite{Ludlow2015},  interferometry pulses can be performed by driving a Rabi oscillation with a low-linewidth laser frequency stabilized to an optical cavity, allowing to build an AI following similar sequences as for two-photon transitions \cite{Hu2017,Hu2020,Rudolph2020}.

In a single-photon Mach-Zehnder interferometer the shot-noise-limited acceleration sensitivity is given by $\delta a = (nk\sqrt{N}T^{2}1)^{-1}$, where $k=2\pi/\lambda$ is now the wave number of the clock laser of wavelength $\lambda$. To increase the sensitivity, one can adopt large momentum transfer (LMT) techniques resulting in increasing $n$ \cite{Rudolph2020}. Due to the small dipole moment of the clock transition, high-intensity laser beams (1 kW/cm$^2$) are generally needed to obtain a sufficiently high Rabi frequency. Moreover, a high-spectral purity laser source is necessary to drive a high quality factor $Q$ optical transition \cite{Hu2020, Ludlow2015}. 

With this in mind, various concepts of space GW detectors have been proposed \cite{ElNeaj2020,Tino2019b}, first relying on the extension of previous space gradiometry concepts \cite{Hogan2011}, and finally raising the idea that there is little conceptual difference between the atom interferometry based differential phase readout and time delay heterodyne measurements \cite{Hogan2016,Norcia2017}. This led to concepts for space mission relying on optical clocks \cite{Kolkowitz2016,Su2018,Takamoto2020} as pictured in Fig. \ref{fig:spaceground}.

Single photon AIs are also central to ground based projects currently under development \cite{adamson_proposal_2018,Badurina2020}, where the key idea is to exploit the laser frequency noise immunity to develop single arm GW detectors that could extend their sensitivity and open new fields such as Dark Matter or Dark Energy surveys. While clock transitions can open prospect of a major leap in sensitivity, and might allow for using very long resonant cavities, there are still many open questions about the noise-induced hindering of GW signatures that are shared with the current designs relying on Raman transitions \cite{Canuel2020}.

\subsection{Noise sources \label{sec:noise}}

In this section, we will consider the main noise sources related to the geometry and functioning principle of an AI GW detector. Reaching a given GW target sensitivity will require to consider other backgrounds impacting the sensitivity, which are discussed in detail in~\cite{ELGARtecno2020}.

These main noise sources can be understood by analysing the standard geometry introduced earlier, i.e. the gradiometric configuration based on the interrogation of distant atoms clouds using two-photon transitions as shown in Fig.~\ref{fig:AIgradio}. In this setup, the AIs are created by pulsing a common interrogation laser retro-reflected to obtain the two counter-propagating electromagnetic fields that drive the two photon transitions. 
At the time of each pulse, the difference of the laser phase $\Delta\varphi_{las}$ between the two interrogation fields is imprinted on the atomic phase. At the output the interferometer, the measured variation of atom phase $\Delta \phi$ results from the fluctuation of the interrogation laser phase during the duration of the interferometer~\cite{Antoine_2003}; these fluctuations may arise from a GW signal or from different noise sources affecting the laser phase. The specific number and sequence of pulses used to manipulate the atomic wavefunction define the sensitivity of the measurement. A given AI geometry is thus associated with a sensitivity function $g(t)$~\cite{Cheinet08} which provides the interferometer output $\Delta \phi$ as a function of $\Delta\varphi_{las}$:

\begin{equation}\label{eq:sensibfunction}
\Delta \phi(X_i,t)=n \int^{\infty}_{-\infty}\Delta\varphi_{las}(X_i,\tau)g'(\tau-t)d\tau+\epsilon(X_i,t) \, ,
\end{equation}
where $2n$ is the number of photons coherently exchanged during the interrogation process and $\epsilon(X_i,t)$ is the detection noise, i.e. the atom shot noise~\cite{Itano1993} that we now further explain.

At the output of the interferometer, the transition probability $P$ between the two atom states coupled by the interrogation process is given  by a two-waves interference formula $P=1/2 \left[1-\cos (\Delta \phi) \right]$, and the transition probability is usually measured by fluorescence of the atomic ensemble to recover the atomic phase. During this measurement process, the wavefunction of each atom is projected on one of the two states with a probability respectively of $P$ and $1-P$. The noise in the evaluation of P, and thus $\Delta \phi$, follows a Poissonian statistics and is inversely proportional to the square root of the number of atoms used for the measurement. As an example, an atom flux of $10^{\text{12}}$~atoms/s will result in an atom shot noise of  $S_\epsilon=\SI{1}{\mu rad/\sqrt{Hz}}$. As seen in Sec.~\ref{sec:sensor2photons}, the use of an entangled source of atom can improve this limitation potentially up to the $1/N$ Heisenberg limit. In the following numerical applications we will consider the same source with 20 dB squeezing \cite{Lucke2011,Hosten2016} that enables to reach a detection limit of $S_\epsilon=\SI{0.1}{\mu rad/\sqrt{Hz}}$.

\begin{figure}[htp]
\centering
\includegraphics[width=.8\linewidth]{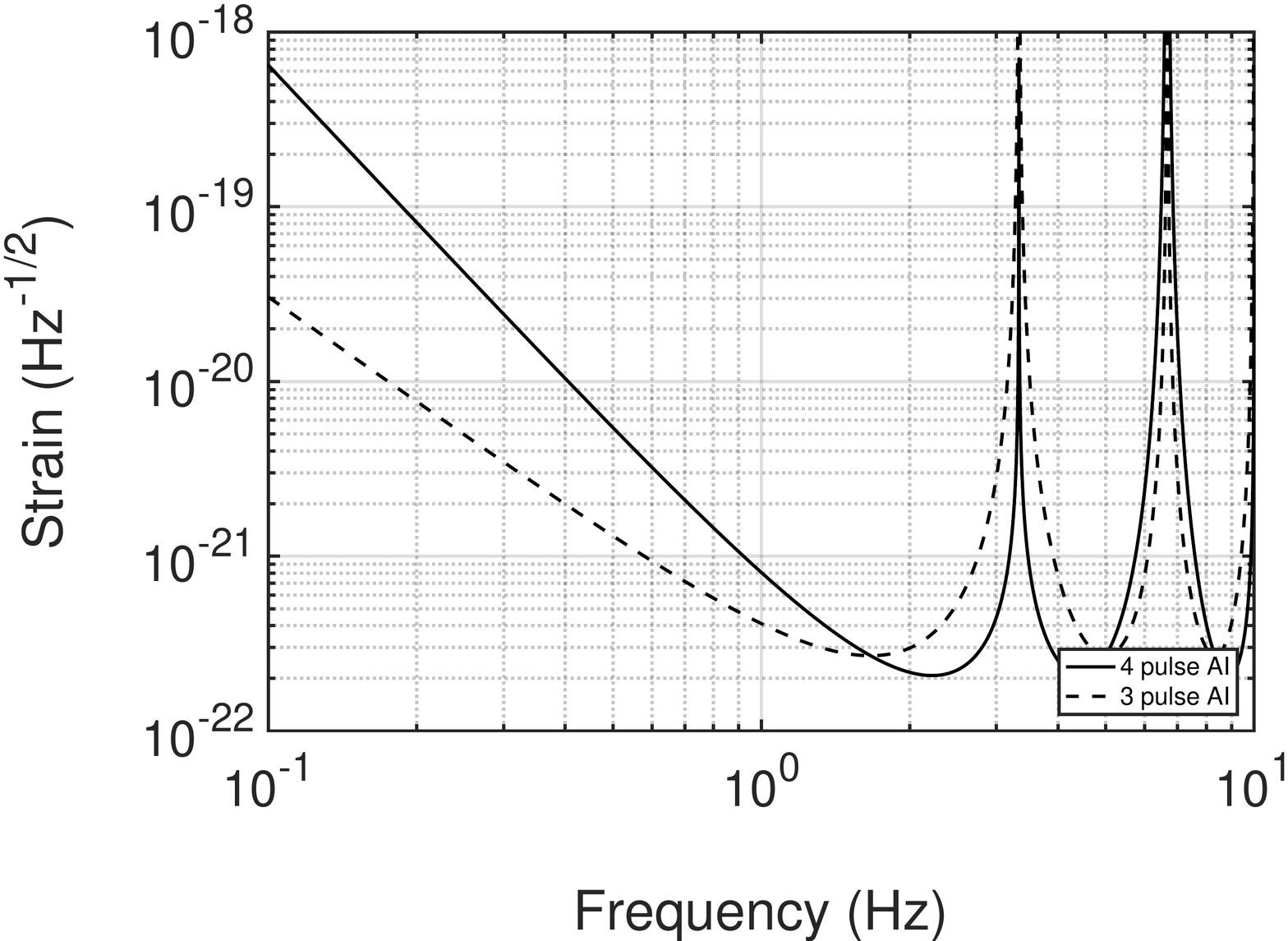}
\caption{GW strain sensitivity of a 16.3~km atom gradiometer for 3-pulses and 4-pulses geometries. Parameters of calculation in the text. \label{fig:sensib}}
\end{figure}

We now derive the GW strain sensitivity of an atom gradiometer: in addition to detection noise, we also consider the main noise sources affecting state-of-the-art GW detectors~\cite{Aasi2015, Acernese2014} such as interrogation laser frequency noise $\delta\nu(\tau)$, vibration of the retro-reflecting mirror $\delta x_{M_X}(\tau)$, and NN that introduces fluctuations of the mean atomic trajectory along the laser beam direction $\delta x_{at}(X_i,\tau)$. These effects affect the local variation of $\Delta\varphi_{las}$ and Eq.~(\ref{eq:sensibfunction}) can be written ~\cite{Canuel2020}:
\begin{align}\label{eq:Deltaphix}
\Delta \phi(X_i,t)&=\int^{\infty}_{-\infty}2nk_l\Big[\Big(\frac{\delta\nu(\tau)}{\nu}+\frac{h(\tau)}{2}\Big)(M_X-X_i)\nonumber\\
&+\Big[\delta x_{M_X} (\tau)-\frac{M_X-X_{i}}{c}\delta x'_{M_X}(\tau)\Big]-\delta x_{at}(X_i,\tau)\Big] g'(\tau-t) d\tau \nonumber\\
&+\epsilon(X_i,t) \, ,
\end{align}
where $h(\tau)$ is the GW strain variation and ${k_l=\frac{2\pi\nu}{c}}$ is the wave number of the interrogation laser. By considering the gradiometric signal $\psi(X_i,X_j,t)$ given by the difference of the outputs of the two AIs placed at positions $X_i$ and $X_j$, we obtain:
\begin{align}\label{eq:diffatphi}
\psi(X_i,X_j,t)=\Delta \phi(X_i,t)-\Delta \phi(X_j,t)\nonumber=\int^{\infty}_{-\infty}2n k_l\Big[\Big(\frac{\delta\nu(\tau)}{\nu}+\frac{h(\tau)}{2} - \frac{\delta x'_{M_X}}{c}\Big)L \nonumber\\
+\delta x_{at}(X_j,\tau)-\delta x_{at}(X_i,\tau)\Big]g'(\tau-t) d\tau +\epsilon(X_i,t)-\epsilon(X_j,t) \, ,
\end{align}
where $L=X_j-X_i$ is the gradiometer baseline. Considering that the detection noise of the two AIs is uncorrelated, the PSD of this signal is then given by:
\begin{align}\label{eq:Spsi}
S_{\psi}(\omega)=(2nk_l)^2 \Big[\Big(\frac{S_{\delta \nu}(\omega)}{\nu^{2}}+\frac{S_h(\omega)}{4}+\frac{\omega^{2}}{c^{2}}S_{\delta x_{M_X}}(\omega)\Big)L^2\nonumber \\
+S_{NN_1}(\omega)\Big]|\omega G(\omega)|^2+2 S_{\epsilon}(\omega) \, ,
\end{align}
where $S_.$ denotes the Power Spectral Density (PSD) operator, $G$ is the Fourier transform of the sensitivity function $g$ and $S_{NN_1}$ is the PSD of the relative displacement of the atom test masses due to Newtonian noise: $NN_1(t)=\delta x_{at}(X_j,t)-\delta x_{at}(X_i,t)$.
Using this gradiometric configuration, the signal-to-noise ratio (SNR) for detecting GWs at a given frequency is given by dividing the GW term of Eq.~(\ref{eq:Spsi}) by the sum of all the other terms.
The strain sensitivity at a given frequency is then obtained by considering the GW strain corresponding to an SNR of 1, and is given by:
\begin{align}\label{eq:Sh}
S_{h}&=\frac{4S_{\delta \nu}(\omega)}{\nu^{2}}+\frac{4S_{NN_1}(\omega)}{L^2}+\frac{4\omega^{2}S_{\delta x_{M_X}}(\omega)}{c^{2}}+\frac{8S_{\epsilon}(\omega)}{(2nk_l)^2L^{2}|\omega G(\omega)|^{2}}.
\end{align}
The detector should be ideally designed such that the dominant noise is the detection limit coming from atom shot noise, last term of Eq.~(\ref{eq:Sh}). This term is depending on the transfer function of the AI, which must be chosen mainly with respect to sensitivity to spurious effects and compatibility with LMT schemes. As an example, we report on Fig.~\ref{fig:sensib} the strain sensitivity curves for standard 3-pulses ``$\pi/2$-$\pi$-$\pi/2$" and 4-pulses ``$\pi/2$-$\pi$-$\pi$-$\pi/2$" interferometers of respective sensitivity transfer functions $|\omega G_{3p}(\omega)|^{2}=16 \sin^{4}\left(\frac{\omega T}{2}\right)$ and $|\omega G_{4p}(\omega)|^{2}=64 \sin^2\left(\omega T\right)\sin^{4}\left(\frac{\omega T}{2}\right)$  \cite{Cheinet08,leveque}. 
The curves are calculated for a total interferometer time of $2T= 600$~ms, a number of  photon transfer $n=1000$, a shot noise level of $S_\epsilon=\SI{0.1}{\mu rad/\sqrt{Hz}}$ and a gradiometer baseline of $L=16.3$~km. These parameters are considered for the matter wave based GW detector proposed in~\cite{Chaibi2016}, see Sec.~\ref{sec:GGN}. We observe that the strain sensitivity reaches respectively for the two interferometer configurations about $2.1 \times 10^{-22}/\sqrt{\rm{Hz}}$ and $2.7 \times 10^{-22}/\sqrt{\rm{Hz}}$ at a corner frequency (lowest frequency at which the maximum sensitivity of the detector is reached) of 1.7 and 2.2 Hz. For a given AI geometry, the corner frequency, and thus the detection bandwidth, is inversely proportional the total interferometer time.

We now discuss the other noise terms listed in Eq.~(\ref{eq:Sh}). For the seismic noise contribution, we observe that the position noise of the retro-reflecting mirror $\delta x_{M_X}$ (see Eq.~(\ref{eq:Deltaphix})) is a common noise for the two AIs, and is thus rejected in the gradiometer signal. Still, seismic noise can impact the measurement through residual sensitivity to mirror velocity and induce a strain limitation of $4\omega^{2}/c^{2}S_{\delta x_{M_X}}(\omega)$ (third term of Eq.~(\ref{eq:Sh})). As an example, considering a seismic displacement noise at 1~Hz of $10^{-9}~\rm{m}/\sqrt{\rm{Hz}}$, typical of sites with good seismic conditions, the strain limitation will be $2\times10^{-17}/\sqrt{\rm{Hz}}$. An atom gradiometer with parameters of Fig.~\ref{fig:sensib} would therefore require a suspension system to reach the atom shot noise limit. Remarkably, such system would have less stringent requirements with respect to the suspensions needed in an optical GW detector ~\cite{Aston_2012,ACERNESE2010182}: considering for example the case of the 4 pulse AI of Fig.~\ref{fig:sensib}, an isolation factor of only $2.5\times10^{4}~$ is needed at a frequency of 1 Hz. This difference stems from the important common mode rejection factor introduced by the gradiometric configuration.

For what concerns frequency noise of the interrogation laser, we observe that the relative fluctuations it causes are indistinguishable from the GW effect. Examining again the example of Fig.~\ref{fig:sensib}, the required laser stability in order to be be shot noise limited is about five orders of magnitude beyond the state-of-the art of pre-stabilized lasers~\cite{Robinson2019}. This issue can be solved adopting, as in optical interferometry, a two orthogonal arm configuration, see later Sec.~\ref{ssec:elgar}.

Local gravity perturbations of different geophysical origins, commonly referred to as Newtonian noise, create spurious gradiometric signals. Indeed, the atom gradiometer configuration is similar with an optical GW detector with the atoms as test masses instead of mirrors. Thus, any differential gravity perturbations between the test masses will impact the signal in the same way as GWs do. Since the first generation of GW detectors, NN has been identified~\cite{Saulson1984} as a important source of noise and extensively studied~\cite{Har2015}. It will constitute a potential limiting factor for third generation detectors such as the Einstein Telescope \cite{Punturo2010} in their low frequency detection window, at a few Hz. As seen in Fig.~\ref{fig:sensib}, atomic gradiometers target a detection bandwidth centered at even lower frequencies and NN represent a critical issue to reach the ultimate detector performances linked to atom shot noise. In the next section, we give NN projections on atom gradiometers and present dedicated methods and detector geometries developed to reduce its impact.

\subsection{Interferometer arrays for rejecting gravity-gradient noise \label{sec:GGN}}

\begin{figure}[htp]
\centering
\includegraphics[width=0.8\linewidth]{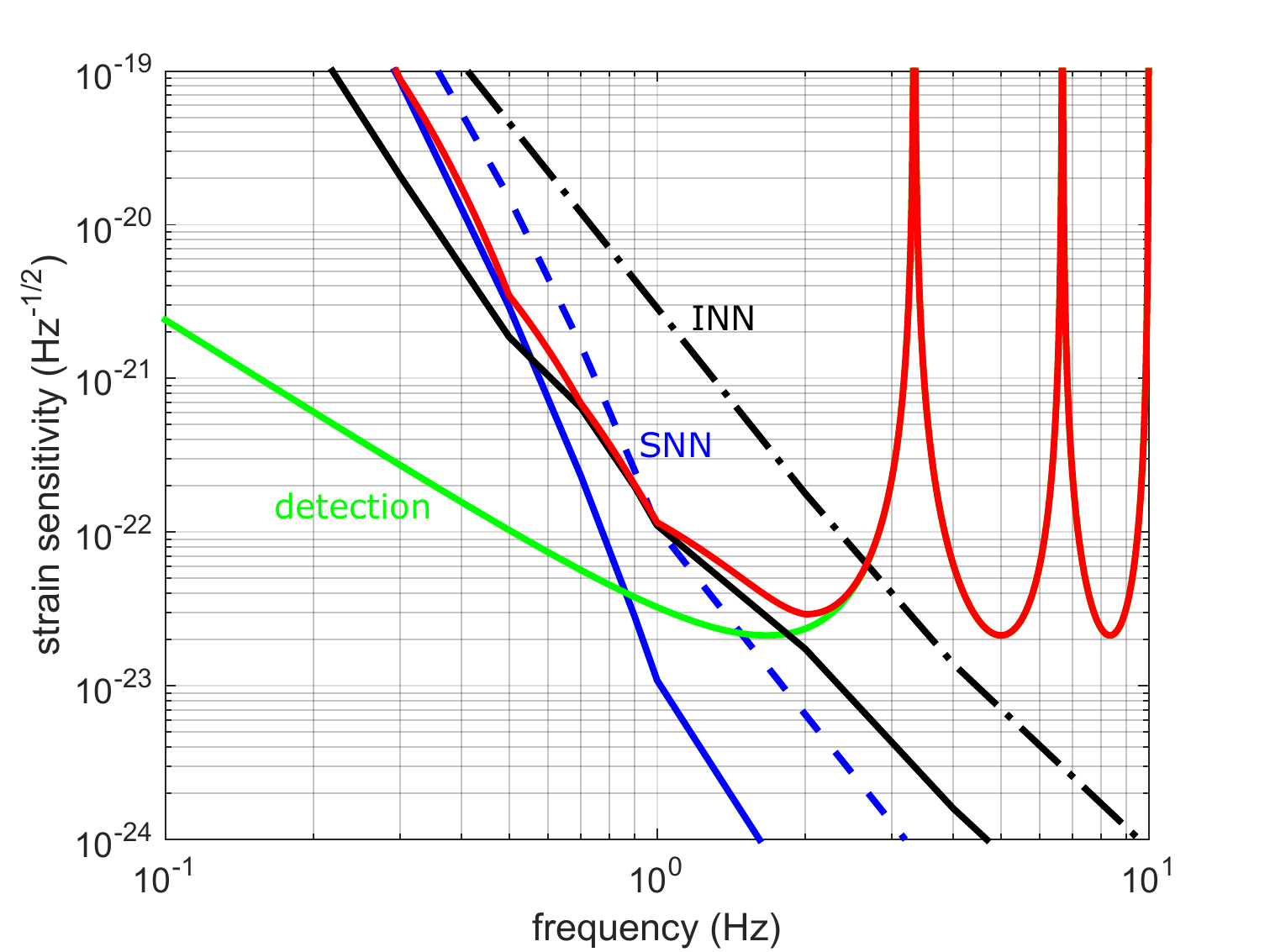}
\caption{Dotted-dashed black (dashed blue): INN (SNN) of a single gradiometer of baseline $L$. Green: atom shot noise strain limitation for a 2D gradiometer array. Black (blue) line: residual strain INN (SNN) using the averaged signal of the array. Red line: overall strain sensitivity of the array. Taken from \cite{Chaibi2016}.
\label{fig:sensitivity_curve}}
\end{figure}

Any anthropological or geophysical process implying a mass transfer or a fluctuation of density of the medium around the detector can be a source of NN. Among these different sources, the main stationary components early identified as possible limitations for GW detector in the infrasound domain come from medium density fluctuations due to the local atmospheric and seismic activity~\cite{Saulson1984}, respectively named Infrasound and Seismic Newtonian noise (INN and SNN). Fig.~\ref{fig:sensitivity_curve}~\cite{Chaibi2016} shows projections of both contributions on the strain measurement of a single gradiometer with a baseline of $L=16.3$~km (dashed black and blue curves). The density variations are calculated from ~\cite{Harms2013,Saulson1984} using as input an air pressure fluctuation spectrum of $\Delta p^2(\omega)=0.3 \times 10^{-5}/(f/1\text{Hz})^2\ \text{Pa}^2/\text{Hz}$ and a seismic noise of $1\times 10^{-17} \ \text{m}^{2}\text{s}^{-4}/\text{Hz}$ at 1 Hz. We observe that for frequencies $<1$ Hz both noise curves stand well above the shot noise limit of a single gradiometer as discussed for Fig.~\ref{fig:sensib}.  

\begin{figure}[htp]
\centering
\includegraphics[width=0.7\textwidth]{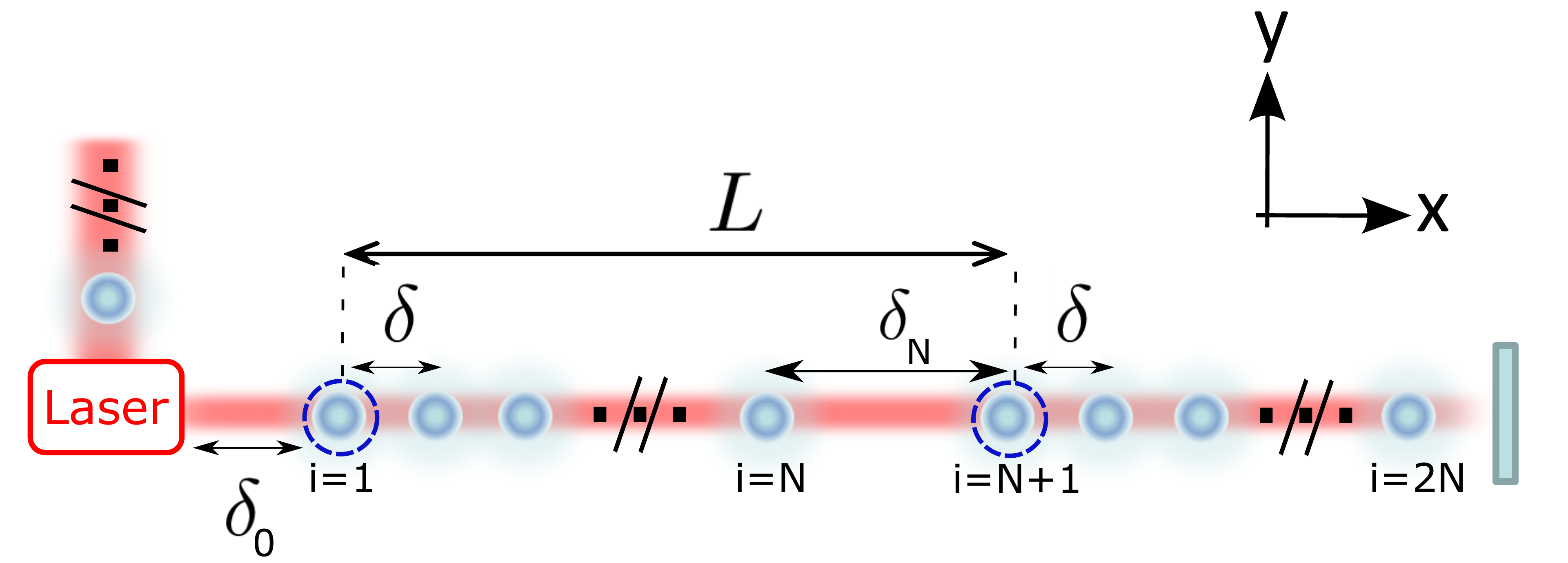}
\caption{Proposed detector geometry, based on a 2D array of atom gradiometers of baseline $L$ regularly separated by a distance $\delta$. Taken from \cite{Chaibi2016}.}
\label{fig:schematic_detector}
\end{figure}

To reduce the impact of NN on atom interferometry based GW detectors, it has been proposed to use not a single gradiometer but an array of them \cite{Chaibi2016}, with a geometry optimised to statistically average the NN. Indeed, GWs and NN signals will have a different spatial signature over the gradiometers of the array: while GWs have a strict plane wave structure, NN has a coherence length of a few kilometers typically in the infrasound domain for the sources considered. We now detail further the method and performances obtained using the detector geometry of Ref.~\cite{Chaibi2016}, shown in Fig.~\ref{fig:schematic_detector}. 
The detector is formed by two symmetric arms along orthogonal directions, interrogated by a common laser. Each arm is formed by $N=80$ gradiometers of baseline $L=16.3$~km and the separation between the gradiometers is set to $\delta= 200$~m.
Averaging the signals from different gradiometers in each arm averages the contribution of NN but maintains the GW contribution. We therefore consider the difference between the average signals of the gradiometers of each arm, given by: 
\begin{equation}
\label{average}
H_N\left(t\right) =\frac{1}{N}\sum^{N}_{i=1} \psi(X_i,X_{N+i},t)-\psi(Y_i,Y_{N+i},t). 
\end{equation}
As for an optical GW detector, this two arm configuration is immune to frequency noise of the interrogation laser for a $(+)$ polarized GW impinging on the detector. Using the method presented in the previous section and neglecting seismic noise (see details in Ref.~\cite{Canuel2020} for discussion on this term) the strain sensitivity using the average signal of Eq.~(\ref{average}) can be written:
\begin{equation}\label{eq:SHnetwork}
S_h(\omega)=\frac{S_{NN}(\omega)}{L^2}+\frac{4S_{\epsilon}(\omega)}{N(2nk_l)^2L^2|\omega G(\omega)|^2}\, ,
\end{equation}
where  $NN(t)$ is the differential displacement on the test masses of the array induced by NN given by:
\begin{equation}
NN(t)=\frac{1}{N}\sum^{N}_{i=1}\Big[\delta x_{at}(X_{N+i},t)-\delta x_{at}(X_i,t)-\delta x_{at}(Y_{N+i},t)+\delta x_{at}(Y_i,t)\Big]\, .
\end{equation}
Fig.~\ref{fig:sensitivity_curve} shows the residual NN strain limitation of the array $\sqrt{S_{NN}(\omega)}/L$ from INN and SNN (resp. black and blue lines). We observe that both contribution are much reduced with respect to the strain limitations they induce on a single gradiometer (dashed black and blue lines). At 1 Hz, the gain on the INN and SNN is respectively of 30 and 10: the proposed method averages NN with a reduction factor better than $1/\sqrt{N}$ in some frequency range. Indeed, the spacing between gradiometers, set by the parameter $\delta$ is chosen to have anti-correlation of the NN for the two AIs placed at this distance. A variation of delta negligible with respect to the NN correlation length will not spoil the rejection efficiency.

The method presented here opens the way towards NN reduction for ground based GW detector and is now considered in the design of the ELGAR instrument, see Sec.~\ref{ssec:elgar}.

\section{Long baseline atom interferometers}

Long baseline sensors based on the coherent manipulation of matter waves are increasingly considered as disruptive tools to investigate fundamental questions related to the nature of the universe. They propose to drastically advance the scientific knowledge in three specific contests: (i) the search for dark matter \cite{Burrage2015,Hamilton2015,Geraci2016,Sabulsky2019}; (ii) the potential interplay between quantum mechanics and general relativity \cite{Asenbaum2017}; (iii) the detection of GWs in the mid-frequency frequency band uncovered by the LISA and LIGO-Virgo interferometers.

The size of the experimental apparatus has direct implications on the ultimate sensitivity curve of atom gravity-gradiometers, i.e. the configuration commonly adopted to measure tiny variations of the gravitational field: increasing the distance between the two atomic sensors improves linearly the instrument sensitivity to strain, given that the baselines typically considered are much shorter than the GW wavelength. Allowing for a wider wavefunction separation has two potential outcomes: (i) exploited to increase the interrogation time, it shifts the sensitivity curve to lower frequencies; (ii) used to transfer a larger momentum separation with LMT techniques shifts the sensitivity curve vertically thus improving it.

All these configurations benefit from a longer baseline, and justify the recent trend to build longer and longer instruments. The first important leap in the size of AIs is represented by the ~10 m tall atomic fountains realized about 10 years ago, like those in Stanford \cite{Dickerson2013}, and Wuhan \cite{Zhou2011}. Other instruments of similar size are being developed nowadays both with the coherent manipulation along the vertical direction (e.g. Hanover \cite{Hartwig2015} and Florence) or the horizontal one (e.g. MIGA prototype).

We are now well into the second dramatic increase of the instrument size, with several experiments being built or proposed to realize on Earth AIs with baselines from a few hundred meters to a few tens of kilometers. At the same time several actions are being carried out to study the potential scientific outcome of a future long baseline instrument operated in space; different aspects - like specific technical solutions, instrument configurations, orbit selection, measurement protocols, and target signals - are being investigated for measurement baselines ranging from a few km to several millions of km.

Within the actual initiatives to realize long baseline atom gravity~gradiometers on Earth we can mention the following ones:

\begin{itemize}
    \item the Matter-wave laser Interferometric Gravitation Antenna (MIGA) experiment \cite{Canuel2018} (see Sec. \ref{ssec:miga}), a horizontal interferometer that is being realized in the underground environment of the ``Laboratoire Souterrain {\`a} Bas Bruit" (LSBB)~\cite{gaffet2019}. Cold rubidium atomic ensembles launched in free-fall and coherently manipulated with two horizontal, vertically displaced, and 150 m long laser beams will measure the differential gravity acceleration between the two extremes of the setup. The first generation instrument will reach a strain sensitivity of 2$\times$10$^{-13}$ Hz$^{-1/2}$, and will be a demonstrator for instruments of a later generation, characterized by an improved sensitivity thanks to more advance atomic manipulation techniques, a longer baseline, and better protocols to reduce GGN.
    
    \item The Mid-band Atomic Gravitational Wave Interferometric Sensor (MAGIS) experiment \cite{Graham2017,Coleman2019} developed by a US consortium (see Sec. \ref{ssec:magis}); it plans a three phases experimental activity on Earth to prepare a future space mission for an instrument capable of detecting GWs in the frequency band [30 mHz--10 Hz]. The first two phases are already being developed, and they consists, respectively, in a 10 m atomic fountain in Stanford (CA-US) and in a 100 m vertical detector being realized in an existing vertical shaft at Fermilab (IL-US). The third detector has been proposed with a vertical baseline of 1 km and could be located at the Sanford Underground Research Facility (SURF in SD-US). The common solutions pursued for the three preparatory phases are the vertical interrogation configuration, which naturally opens towards long interrogation intervals, and the adoption of ultra-cold strontium atoms as gravitational probe.
    
    \item The Zhaoshan long-baseline Atom Interferometer Gravitation Antenna (ZAIGA) experiment in Wuhan (China) \cite{Zhan2019} (see Sec. \ref{ssec:zaiga}), which consists in an underground facility for experimental research on gravitation, and which will also host an horizontal, three 1 km arms on a triangle GW detector using rubidium atoms.

    \item The Atom Interferometry Observatory and Network (AION) experimental program (see Sec. \ref{ssec:aion}), consisting of 4 successive phases to realize successively a 10 m, 100 m, 1 km and a satellite mission targeting dark matter search and GW detection in the mid-frequency band. AION proposes also to be operated in a network configuration with other GW detectors to optimize the scientific output by implementing multiband GW astronomy techniques \cite{Sesana2016} or exploiting the uncorrelation of far located instruments to look for stochastic background \cite{Allen1999}.  

    \item The Very Long Baseline Atom Interferometer (VLBAI) experiment (see Sec. \ref{ssec:vlbai}), 10 m long atomic fountain, where ultra-cold ytterbium and rubidium atoms will be used to test several pillars of quantum mechanics and general relativity, looking for the intrinsic nature of the decoherence mechanisms, violation of the equivalence principle, and developing at the same time the enhanced atom interferometry tools to achieve the sensitivity required for detecting GWs. \medskip

\end{itemize}

Other projects are in their very preliminary study phase, as is the case of the Italian project MAGIA-Advanced, which is studying the feasibility of a vertical instrument with a baseline of a few 100s meters to be installed in a former mine shaft in Sardinia \cite{Canuel2020}, or are at the stage of study proposals such as the The European Laboratory for Gravitation and Atom-interferometric Research (ELGAR) \cite{Canuel2020,ELGARtecno2020} and the Atomic Experiment for Dark Matter
and Gravity Exploration in Space (AEDGE) \cite{ElNeaj2020}. ELGAR, described in depth in Sec. \ref{ssec:elgar}, proposes an underground array of gravity gradiometers with a total baseline of 32 km, adopting advanced atom interferometry techniques to mitigate Newtonian noise and achieve the required sensitivity to detect GWs in the [0.1--10 Hz] frequency band with a terrestrial instrument. AEDGE studies different space configurations exploiting matter wave sensors to push the boundaries of fundamental science, most notably concerning the nature of dark matter and the search of GWs in the frequency band intermediate to the maximum sensitivity of LISA and LIGO-Virgo.

\subsection{MIGA -- Matter-wave laser Interferometric Gravitation Antenna \label{ssec:miga}}

The MIGA antenna~\cite{Canuel2018} is a French ANR funded “Equipement d’Excellence” project to build a large scale infrastructure based on quantum technologies: a hybrid atom and laser interferometer using an array of atom sensors to simultaneously measure the Gravitational Waves (GWs) and the inertial effects acting in an optical cavity. The instrument baseline is designed to reach high sensitivity in the infrasound domain with a peak sensitivity at 2~Hz. This infrastructure will be installed at LSBB~\cite{gaffet2019}, in dedicated galleries located 300~m deep from the surface inside a karstic mountain. This site demonstrates a very low background noise and is situated far from major anthropogenic disturbances determined by cities, motorways, airports or heavy industrial activities.

\begin{figure}[htp]
  \centering
  \includegraphics[width=1\textwidth]{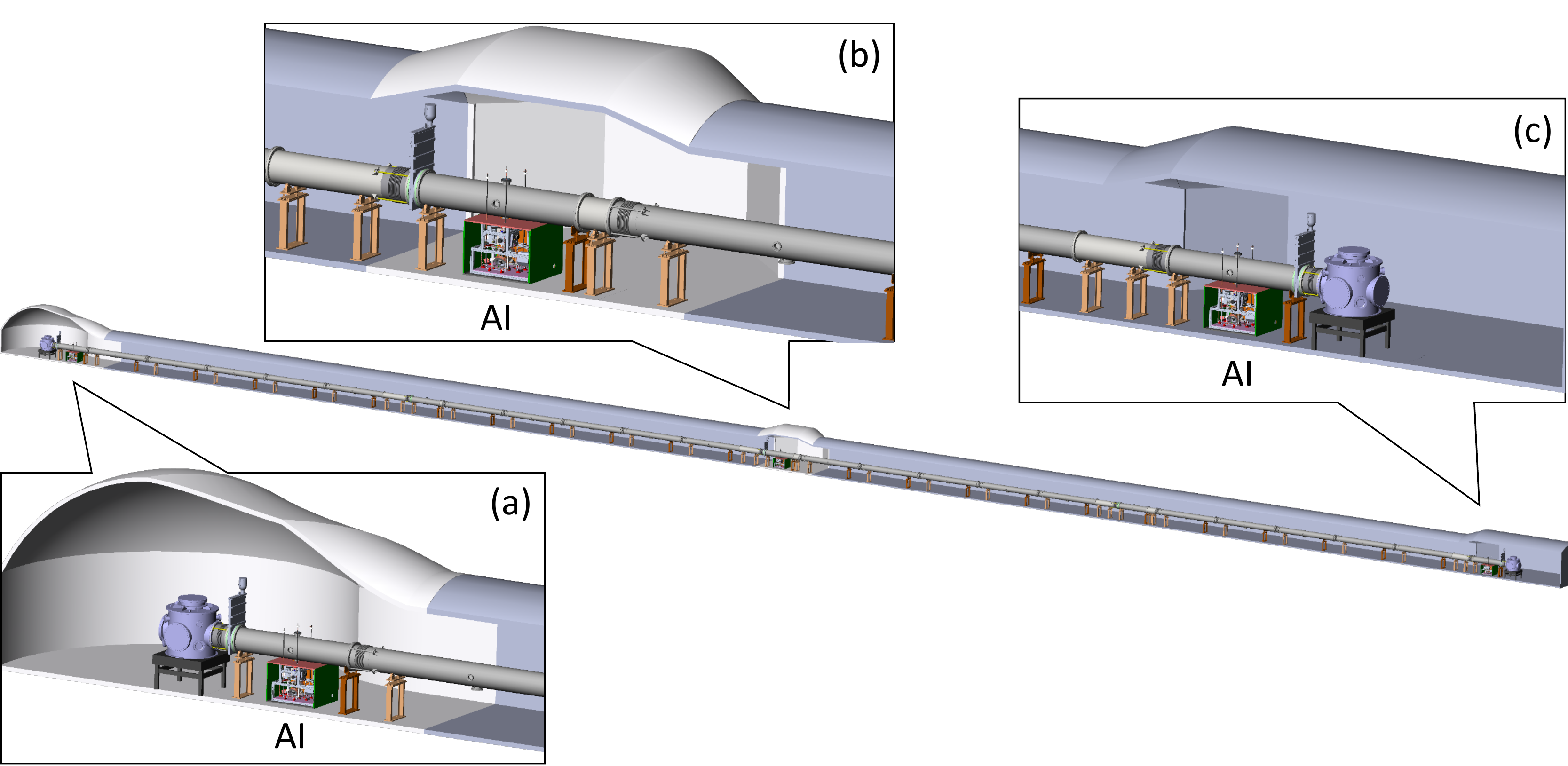}
  \caption{The three AIs of the antenna will be located at (a), (b) and (c). The optical setups for cavity injection will be hosted in room~(a). The vacuum vessel is pre-equipped to add other AIs in the future and reach a total of 5 measurement positions.}\label{LSBB_MIGA}
\end{figure}

A scheme of the antenna at LSBB can be seen in Fig.~\ref{LSBB_MIGA}. Two horizontal cavities are used for the atom interrogation. The optics of the resonators are placed at the extremities of the detector inside vacuum towers (cylinders of 1 m height and diameter shown on panels (a) and (c)). A vacuum vessel with a diameter of 0.5 m and 150~m long hosts the cavity fields that correlate three AIs regularly placed along the antenna (see details (a), (b) and (c) in Fig. \ref{LSBB_MIGA}). The optical benches for the cavity injection are hosted in room (a). In addition to this infrastructure, a shorter version of the instrument, a 6~m cavity gradiometer is in construction at the LP2N laboratory in Talence. This equipment will be used to test advanced atom manipulation techniques that will be implemented later on the antenna.

We now describe the working principle of the antenna by focusing on the measurement process of each AI, detail the status of its construction and give prospects on the scientific results to pursue once the antenna will be operative.

\subsubsection{Functioning and status of the MIGA Antenna}

\begin{figure}[htp]
  \centering
  \includegraphics[width=1\textwidth]{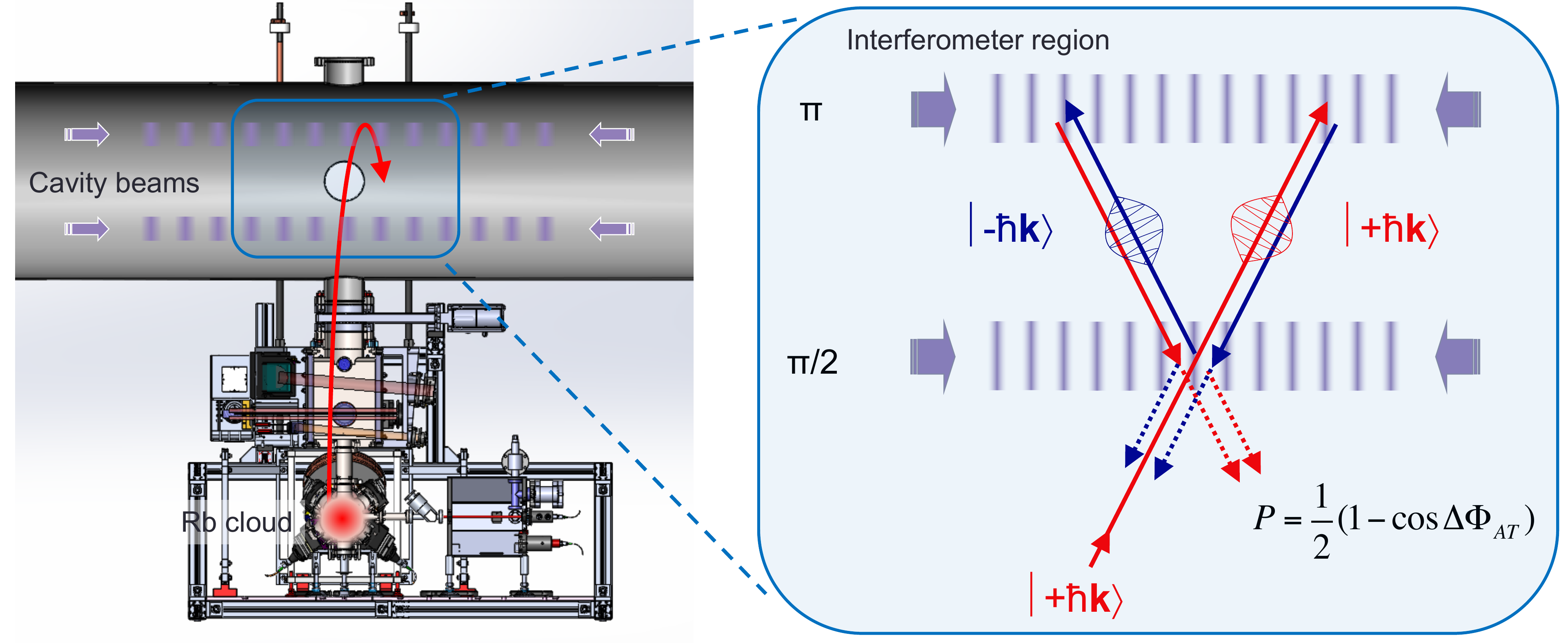}
  \caption{(left)~View of an AI of MIGA. A $^{87}$Rb atom source uses a combination of a 2D and 3D magneto-optic traps. After being trapped and cooled, the atoms are launched on a parabolic trajectory and enter in the interferometric region where they are manipulated using a set of two cavity beams. (right)~Geometry of the 3 pulse AI, taken from \cite{Canuel2018} under CC BY 4.0.}\label{MIGA_AI_view}
\end{figure}

MIGA will require the simultaneous interrogation of the matter wave interferometers placed along its baseline by time-modulation of the laser injected in the cavity. The geometry of each AI, shown in Fig. \ref{MIGA_AI_view}, consists in a 3 pulse interferometer ``$\pi/2$-$\pi$-$\pi/2$".
After a cooling and trapping sequence, a $^{87}$Rb cloud is launched on a vertical parabolic trajectory. When the atom reach the lower cavity beam, they experience a first ``$\pi/2$" Bragg pulse that creates a balanced superposition of the external states $\left |\pm\hbar \mathbf{ k}\right\rangle$. When the cloud reaches the upper cavity beam at the apex of its trajectory, the atoms experience a ``$\pi$" pulse that reverses the horizontal atomic velocity. When the falling atoms reach again the lower cavity beam, a second ``$\pi/2$" pulse closes the interferometer. The state occupancy of the two states at the interferometer output is then measured by fluorescence detection to extract the transition probability and so the interferometric phase $\Delta \phi$.
This signal is determined by the phase difference along the two paths followed by the matter waves inside the interferometer, which is in turn related to the variation of the phase  $\Delta\varphi_{las}$ of the counter-propagating cavity field. The response of the AI can then be obtained from Eq.~(\ref{eq:sensibfunction}) using the sensitivity function $g(t)$ of the three pulse interferometer~\cite{Cheinet08}.

The GW strain variation induced on the phase of the cavity resonating field can be determined with a gradiometric measurement on two atomic sources separated by a distance $L$ along the antenna. Considering the noise sources detailed in Sec.~\ref{sec:noise}, we obtain the following strain sensitivity \cite{Canuel2018}:
\begin{equation}
S_h(\omega)=\frac{4S_{\delta\nu}(\omega)}{\nu^2}+\frac{1}{\left[1+\frac{\omega^2}{\omega_p^2}\right]}\left(\frac{4S_{\mathrm{NN}}(\omega)}{L^2}+\frac{8\omega^2S_x(\omega)}{\omega_p^2L^2}+\frac{8S_{\epsilon}(\omega)}{(2nk_l)^2L^{2}|\omega G(\omega)|^2}\right), 
\label{sensitivity}
\end{equation}
where $G(\omega)$ is the Fourier Transform of $g(t)$, and $\omega_p$ is the cavity frequency pole. When compared to the free-space gradiometric configuration treated in Sec.~\ref{sec:noise}, we observe that the MIGA cavity setup has a similar strain sensitivity for frequencies smaller than the pole of the cavity, whereas it shows a higher sensitivity to seismic noise $S_x(\omega)$, due to the cavity geometry that amplifies the impact of mirror displacement noise on the measurement.

\begin{figure}[htp]
  \centering
  \includegraphics[width=1\textwidth]{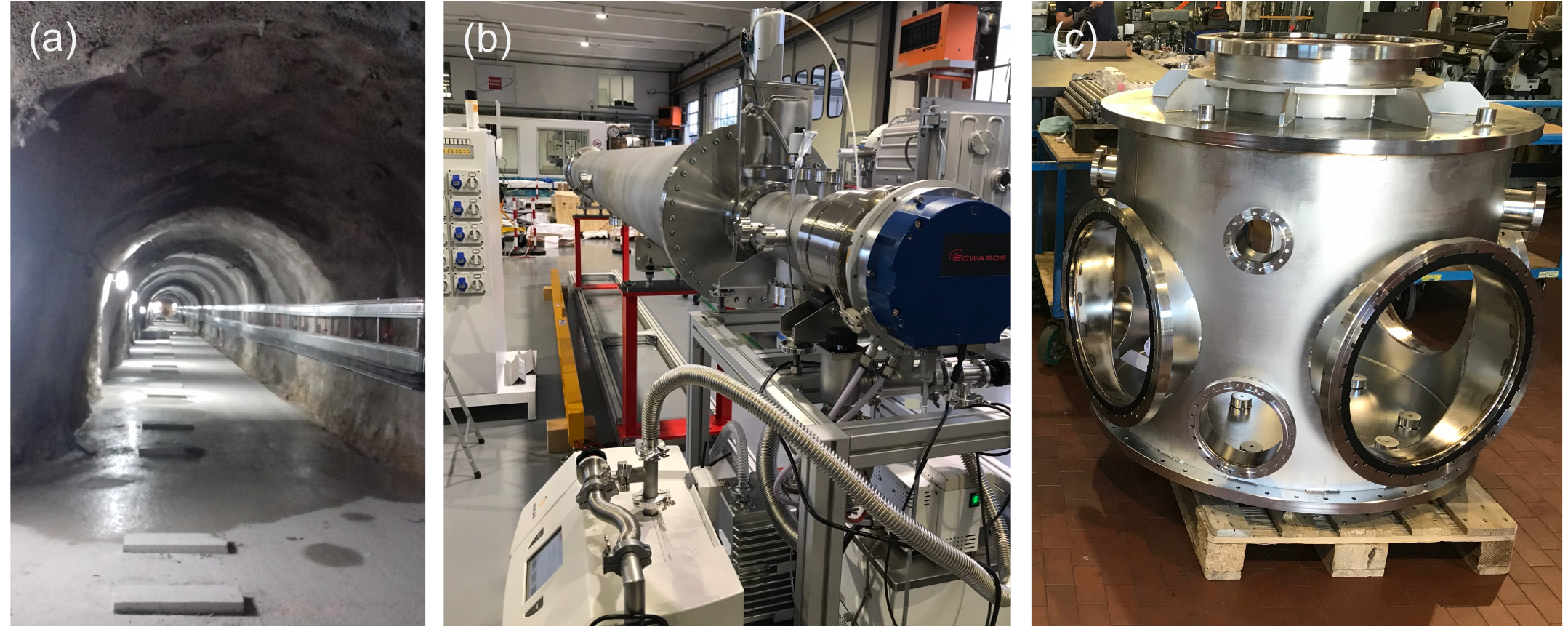}
  \caption{(a) MIGA gallery within LSBB. (b) Standard 6 m long section under vacuum test. (c) Vacuum tower in production at SAES Parma (Italy).}\label{sys_vide2}
\end{figure}

We now describe the status of the realization of the antenna. Starting from 2017, heavy infrastructure works were carried out at LSBB for the installation of MIGA and two new perpendicular 150 m long galleries were bored. These operations lasted till the beginning of 2019. One of the two galleries is being used to install the equipment, the other one will host tests of mass transfer reconstruction and will later allow the instrument upgrade towards a 2D antenna. The MIGA galleries can be seen on Fig.~\ref{sys_vide2}-(a); they have a depth ranging from 300~m to 500~m and their access is at about 800~m from the laboratory main entry. 

\begin{figure}[htp]
  \centering
  \includegraphics[width=0.9\textwidth]{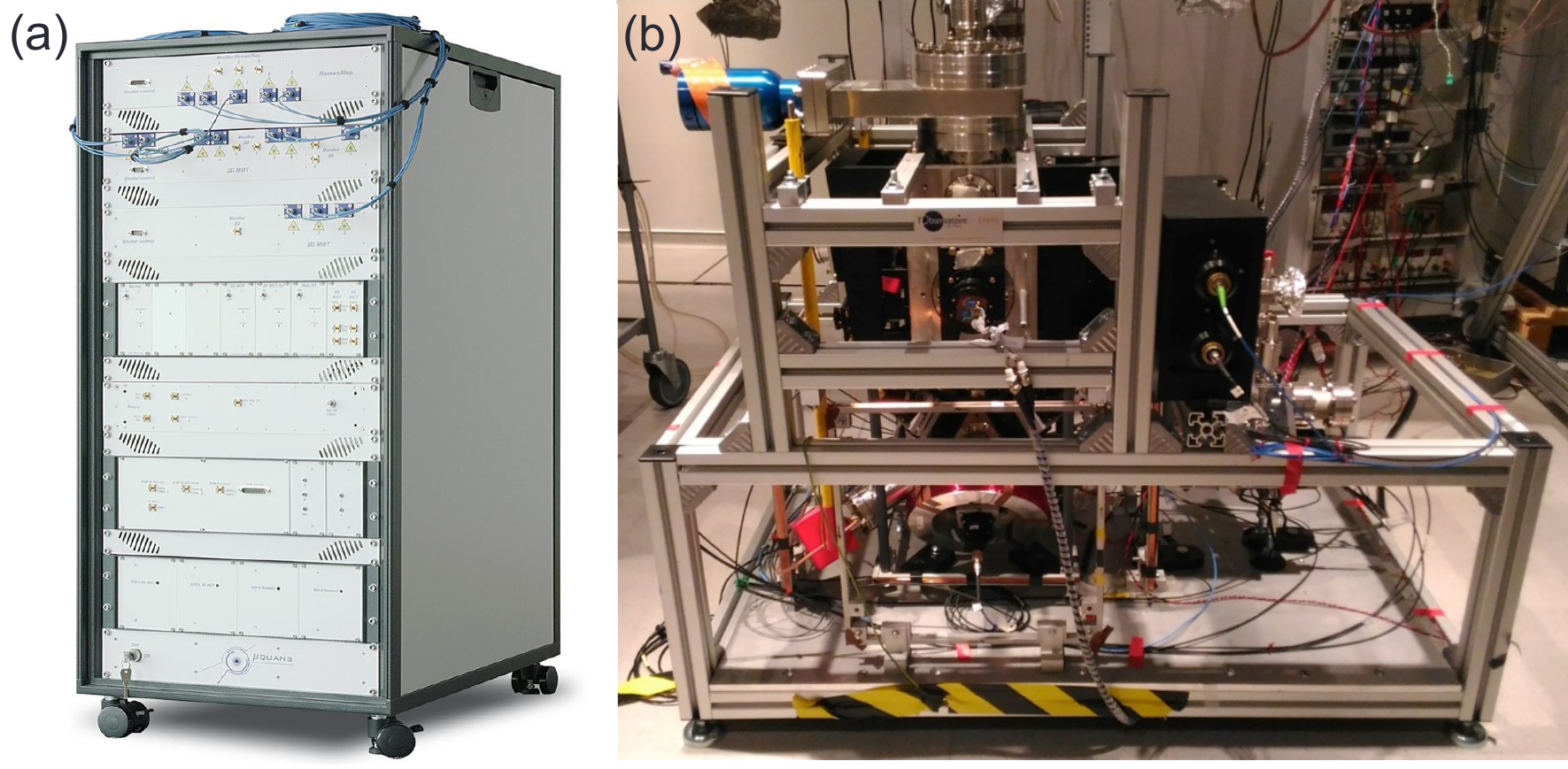}
  \caption{(a) Fiber laser system developed by the Muquans company (image from~\cite{Sabulsky2020}), (b) Cold $^{87}$Rb atom source (image from~\cite{Canuel2018} under CC BY 4.0)}\label{head_laser}
\end{figure}

The vacuum vessel of the antenna is a set of 50~cm diameter SS304 pipes produced by SAES in Parma (Italy) during 2020, see Fig.~\ref{sys_vide2}-(b),(c). It is mainly composed of sections of 6 m long connected using helicoflex gaskets and is designed to reach a residual pressure better than $10^{-9}$~mbar after a baking process up to 200~$^\circ$C. 
A $^{87}$Rb cold atom source for the antenna can be seen on Fig.~\ref{head_laser}-(b).
A combination of 2D and 3D magneto-optic traps prepares about 10$^9$ $^{87}$Rb atoms at a temperature of a few $\mu$K. An optical system based on Raman transitions selects from the cloud a narrow velocity class and a pure magnetic state before entering in the interferometric region. The detection of the atom state is then carried out by fluorescence of the cloud. This system works in combination with a robust and remotely operable fibered laser system~\cite{Sabulsky2020} shown in Fig.~\ref{head_laser}-(a), and a modular hardware control system \cite{Bertoldi2020}.

\subsubsection{MIGA sensitivity and prospects}

\begin{figure}[htb]
\centering
\includegraphics[width=7 cm]{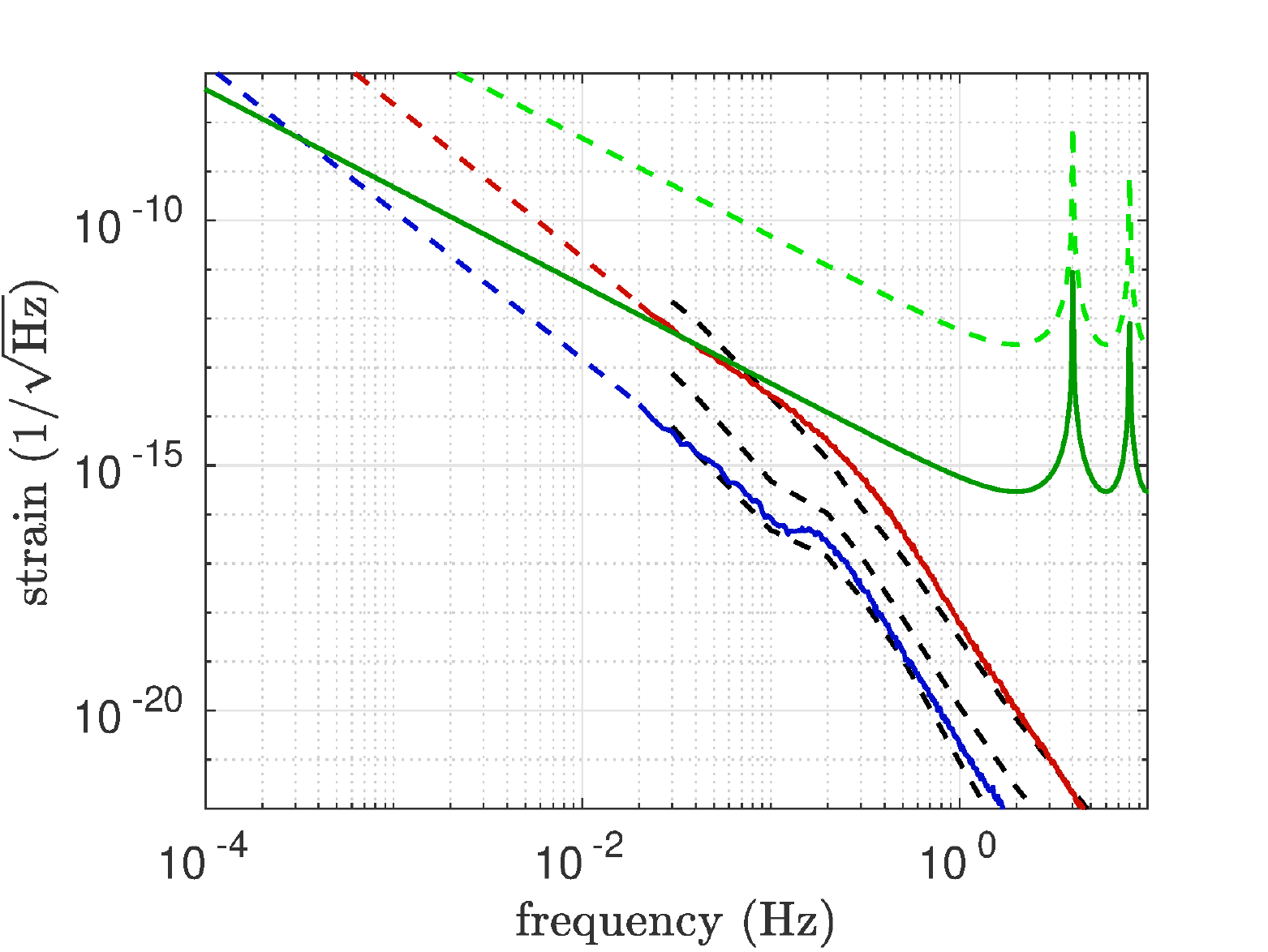}
\caption{Atom shot noise limited sensitivity for the initial (dashed green line) and improved (plain green line) MIGA detector. Projections of strain infrasound GGN using the Bowman atmospheric pressure model~\cite{Bowman2005} (dashed black lines), and using data measured at LSBB for calm (blue) and windy (red) periods. Taken from~\cite{Junca2019}.}
\label{fig:strainMigaINN}
\end{figure}

MIGA aims at an initial stain sensitivity of $2 \cdot 10^{-13}/\sqrt{\mathrm{Hz}}$ at 2~Hz, which is several orders of magnitude short of targeting the GW signals expected in the band. In this sense, MIGA will be a demonstrator for GW detection, and a significant upgrade of both its baseline and the adopted laser and atom optics techniques will be required to fill the sensitivity gap. The initial instrument will be used to study advanced measurement strategies and atom manipulation techniques that could be directly implemented on the experiment so as to improve its sensitivity. These prospects are illustrated in Fig.~\ref{fig:strainMigaINN}, which shows the MIGA strain sensitivity for its initial and upgraded version - the latter with an envisaged use of LMT of $2\times 100$ photon transitions and an improved detection noise of $0.1$~mrad/$\sqrt{\mbox{Hz}}$.

In the short term, the antenna will provide extremely high sensitivity measurements of the local gravity over large baselines which can be used to study how networks of atom gravimeters can resolve the space-time fluctuations of the gravity field. Theses studies are important for future GW interferometers, since GGN, will be a limiting factor for their operation. As an example, we see in Fig.~\ref{fig:strainMigaINN} that infrasound GGN could be detectable in the decihertz range with the upgraded version of the antenna.

\subsection{MAGIS -- Mid-band Atomic Gravitational Wave Interferometric Sensor \label{ssec:magis}}

The MAGIS project \cite{Coleman2019,Graham2017} will realize in different phases very long baseline AIs, with scientific targets ranging from search for dark matter and new forces, to test of quantum mechanics and general relativity at distances not yet investigated. One of the most important objectives is the detection of GWs in the 0.1--10 Hz frequency band, in the sensitivity gap between the Advanced LIGO and LISA experiments.

\begin{figure}[htb]
\centering
\includegraphics[width=11 cm]{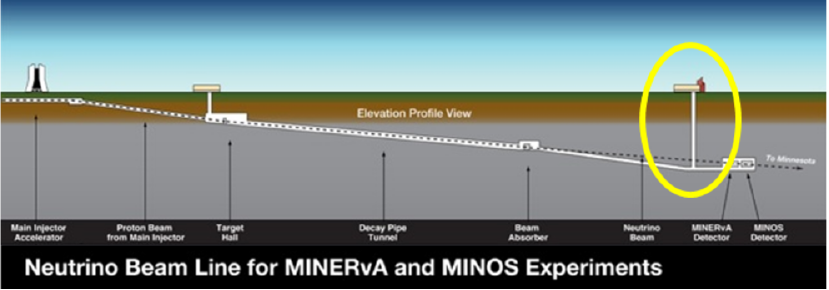}
\caption{Proposed site for the MAGIS-100 experiment, which will exploit the existing NuMI (Neutrino MAin Injector) vertical tunnel at Fermilab, indicated in yellow. From~\cite{Coleman2019}, with permission of T. Kovachy and J. M. Hogan.}
\label{fig:magisConf}
\end{figure}

\begin{figure}[htb]
\centering
\includegraphics[width=9 cm]{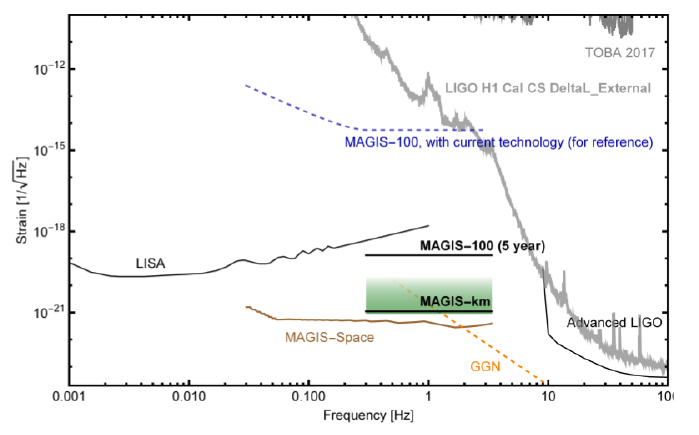}
\caption{GW strain sensitivity curves for different generations and configurations of the MAGIS experiment, both ground- and space based. For comparison purposes are also shown the expected GGN limiting terrestrial detectors, and the sensitivity curves of TOBA \cite{Shoda2017}, Advanced LIGO and LISA. From~\cite{Coleman2019}, with permission of T. Kovachy and J. M. Hogan.}
\label{fig:magisSens}
\end{figure}

The program will have different phases, each based on the development of a gravity gradiometer with increasing length: the first prototype is being already realized in Stanford, and it consists in a 10 m atomic strontium fountain, where key sensitivity enhancement techniques have been already demonstrated, as is the case of LMT clock atom interferometry \cite{Rudolph2020}. The second step will consist in the realization of a vertical, 100 m long atomic fountain, called MAGIS-100, which will exploit an existing vertical shaft at the Fermilab (see Fig. \ref{fig:magisConf}). The planning and construction of the instrument have already started, backed by a consortium formed by Fermilab, Stanford University, Northern Illinois University, Northwestern University, Johns Hopkins University and the University of Liverpool. The long baseline instrument will have several challenging scientific targets, like creating quantum superposition states with unprecedented spatial and momentum separation, search for ultralight dark matter candidates, and test the equivalence principle of general relativity. The primary objective will be however to complement present and future instruments based on optical interferometry in the measurement of GWs, providing high strain sensitivity in the mid-frequency band (see Fig. \ref{fig:magisSens}).

 MAGIS-100 will be the precursor of a terrestrial instrument with an even longer baseline, possible 1 km, for which the SURF laboratory in South Dakota could be a potential candidate site. The final target is a space-based mission \cite{Graham2017}, where the gradiometric measurement uses two atomic sensors on dedicated spacecrafts placed at a distance of 4.4$\times 10^4$ km; the very long baseline, together with the elimination of the disturbance represented by the GGN thanks to the quiet environment obtained with the orbiting satellites will permit to deploy the full potential of atom interferometry. A strain sensitivity $< 10^{-21}$ will be accessible over two frequency decades, also thanks to the possibility to peak the instrument sensitivity at specific frequencies via the ``resonant mode'' configuration \cite{Graham2016}. Even if the space mission is still distant in the future and speculative, preliminary considerations on the instrument error model and on the possible orbit have been studied. The predicted discovery potential will expand manifold what expected with the previous phases of the program, and lists new astrophysical sources like neutron star binaries and intermediate mass black-holes, cosmological sources and stochastic gravitational radiation.

\subsection{ZAIGA -- Zhaoshan long-baseline Atom Interferometer Gravitation Antenna \label{ssec:zaiga}}

\begin{figure}[htb]
\centering
\includegraphics[width=5.5 cm]{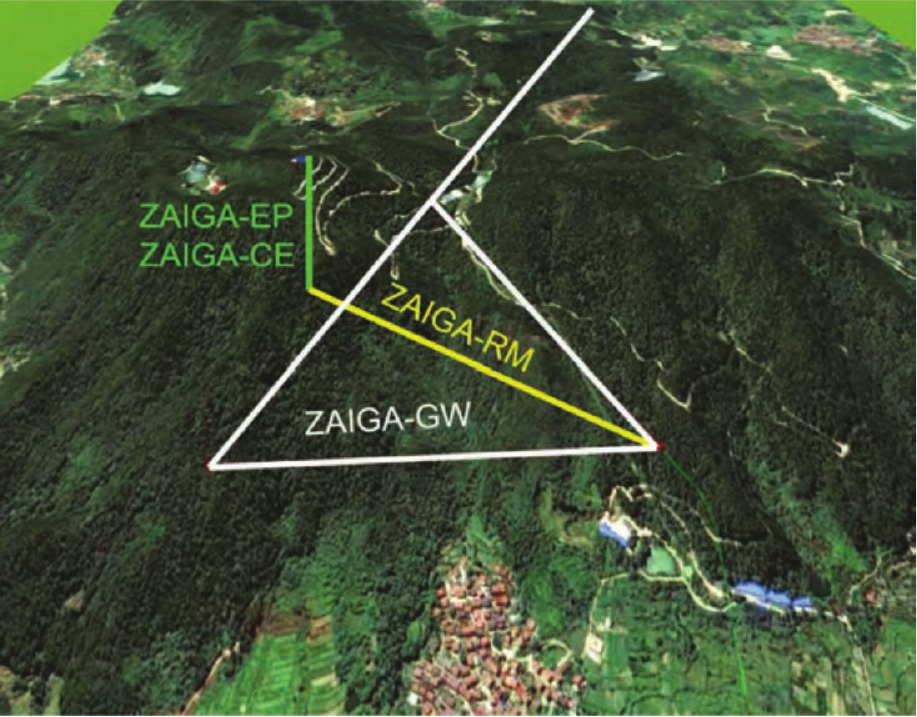}
\includegraphics[width=5.5 cm]{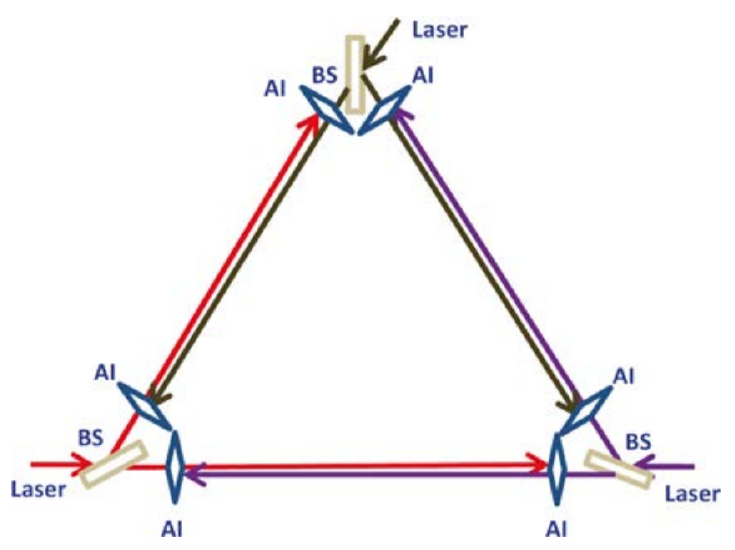}
\caption{(left) Scheme of the underground ZAIGA facility below the Zhaoshan mountain, 80 km south-east of Wuhan (China). The baselines of the different instruments are represented in colors: the 1 km triangular GW detector in white; the gyroscope in yellow; the clock based experiments and the EP test in green. (right) Diagram of the ZAIGA-GW instrument. Republished with permission of World Scientific Publishing Co., Inc., from \cite{Zhan2019}; permission conveyed through Copyright Clearance Center, Inc.}
\label{fig:zaigaConf}
\end{figure}

The ZAIGA underground interferometer facility \cite{Zhan2019} is currently under construction near Wuhan in China (Fig. \ref{fig:zaigaConf}, left). The infrastructure is planned to rely heavily on multiple laser links, and it will implement quantum sensors based on cold atoms for a wide range of purposes: (i) detect GWs in the mid-band using atom interferometry; (ii) test the equivalence principle with a target sensitivity of 10$^{-13}$ or better on the E{\"o}tv{\"o}s parameter; (iii) develop optical clocks for the improved measurement of the gravitational redshift and the realization of a prototype GW detector; (iv) realize an enhanced, long baseline atom gyroscope for improved tests of general relativity.  

The atom interferometry based GW detector, called ZAIGA-GW (Fig. \ref{fig:zaigaConf}, right), will be located 200 m underground to reduce the disturbance due to seismic noise, and it will be developed in two phases: in the first one a triangular configuration with sides of 1 km will be implemented; later the overall size will be increased to 3 km or more. In each side of the structure there will be two counter-propagating laser beams to coherently manipulate the couples of atomic sensors placed along each baseline. The triangular configuration of the instruments gives access to the full strain tensor, which means that both GW polarizations $h_+$ and $h_{\times}$ are measurable.

\begin{figure}[htb]
\centering
\includegraphics[width=8 cm]{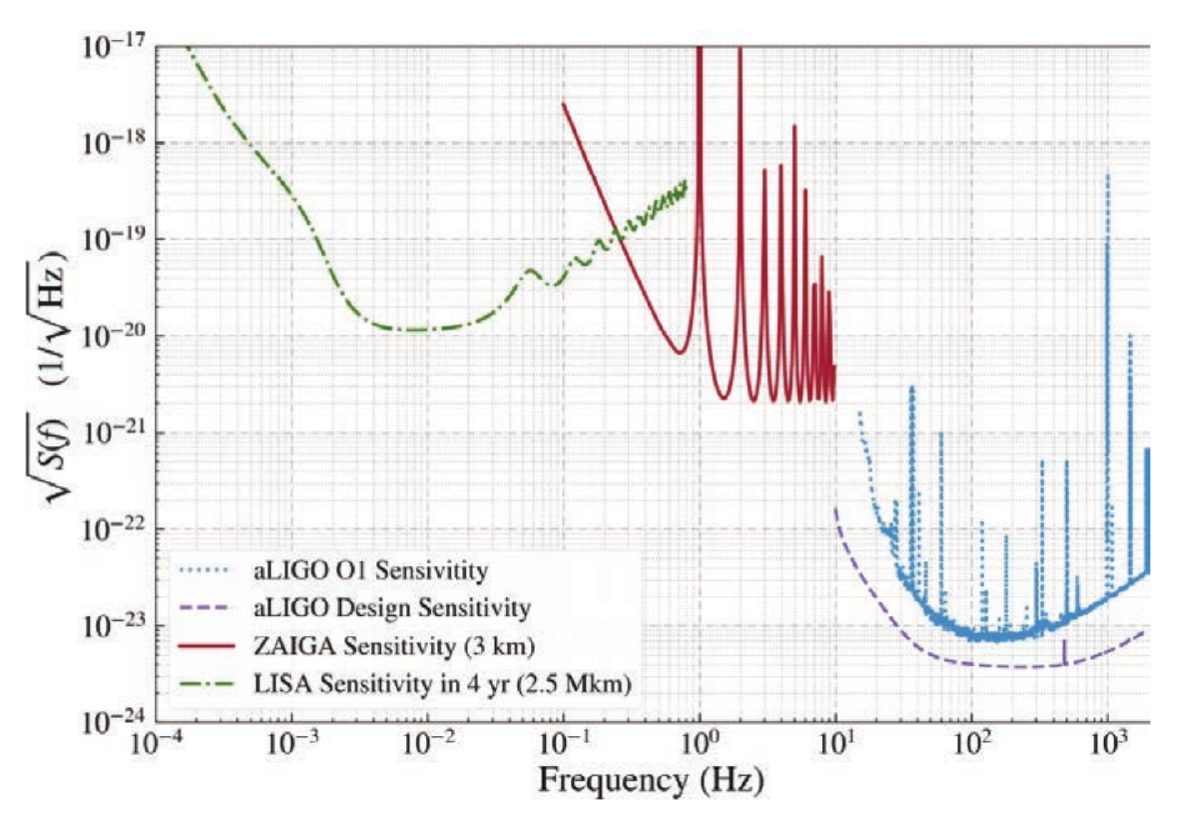}
\caption{Expected strain sensitivity of the 3 km ZAIGA-GW  instrument, compared to the sensitivity curves of LIGO and LISA. Republished with permission of World Scientific Publishing Co., Inc., from \cite{Zhan2019}; permission conveyed through Copyright Clearance Center, Inc.}
\label{fig:zaigaSensitivity}
\end{figure}

The instrument will use cold $^{87}$Rb atoms interrogated in a standard $\pi/2 - \pi - \pi/2$ pulse sequence, and it will target a quantum projection noise limited strain sensitivity below 10$^{-20}$ around 1 Hz, as shown in Fig. \ref{fig:zaigaSensitivity}. To reach such level, high atomic fluxes, the use of sub-shot noise techniques, and adoption of LMT protocols are envisaged. Despite the quiet underground location chosen for its installation, ZAIGA will strongly suffer from Newtonian noise in the detection bandwidth, and mitigation techniques based on vibration isolation and/or suppression via arrays of auxiliary sensors \cite{Chaibi2016} will be pursued.

\subsection{AION -- Atom Interferometry Observatory and Network \label{ssec:aion}}

\begin{figure}[htb]
\centering
\includegraphics[width=4cm]{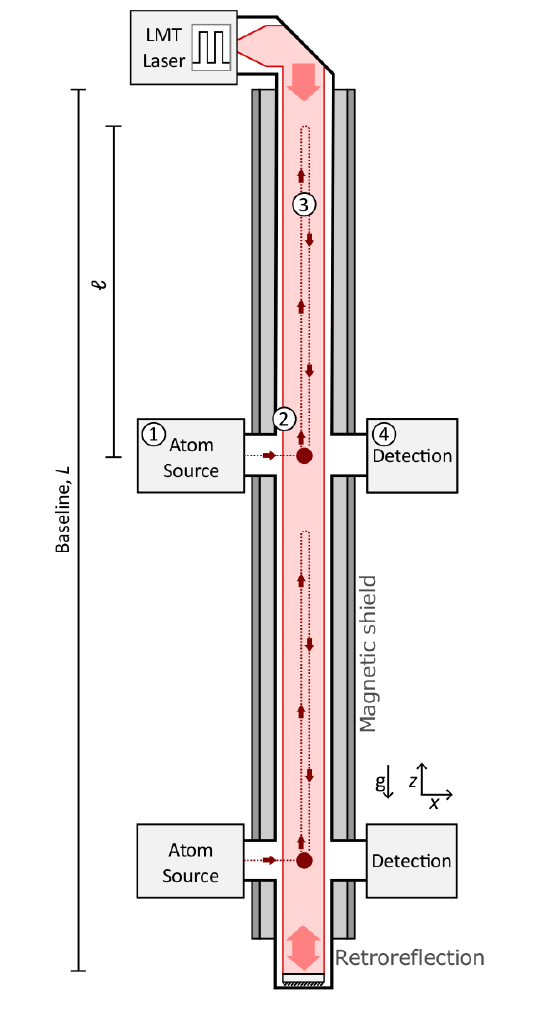}
\raisebox{0.2\height}{\includegraphics[width=7.5cm]{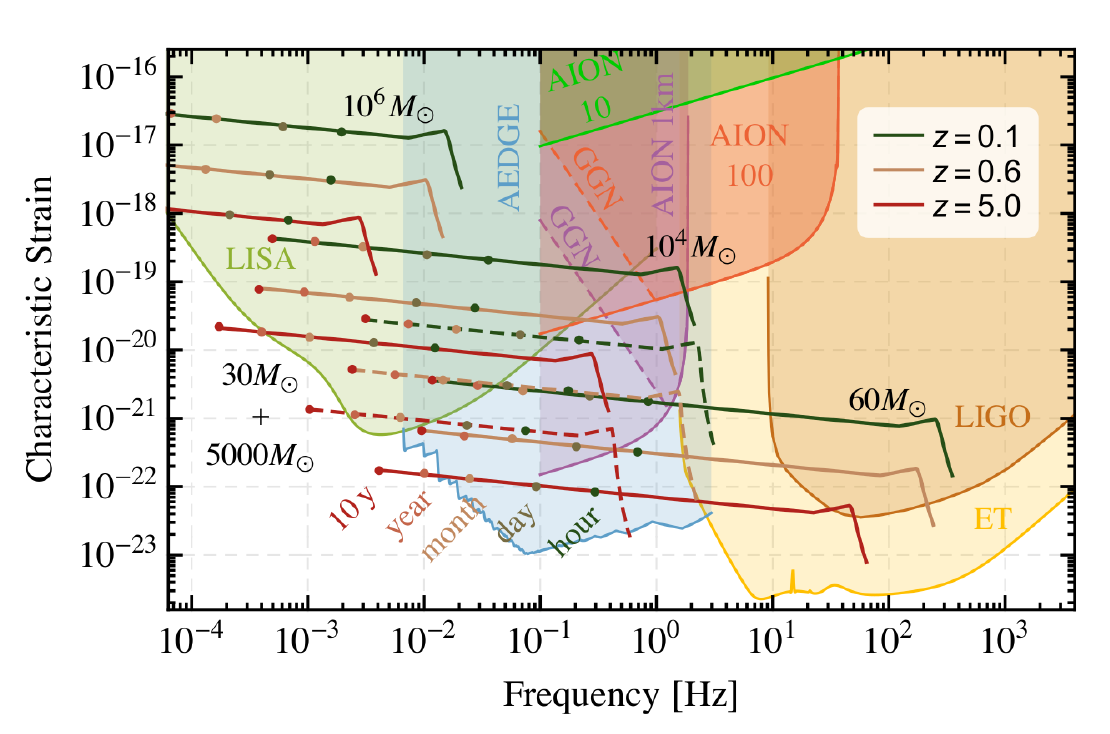}}
\caption{From~\cite{Badurina2020}: (left) Scheme of the AION apparatus, which makes use of two strontium AIs vertically separated and manipulated by a single laser propagating also along the vertical direction. (right) Predicted strain measurement sensitivity for the different AION versions, compared with other existing or planned instruments; the graph shows also the frequency sweeps of binary mergers at various redshift and for different total mass, and the expected gravitational gradient noise for the Earth-based AION detectors. Reprinted from \cite{Badurina2020}, \copyright $\,$ IOP Publishing Ltd and SISSA Medialab Srl, reproduced by permission of IOP Publishing, all rights reserved.}
\label{fig:aion}
\end{figure}

The AION program \cite{Badurina2020}, led by an English consortium, aims to push the boundaries of fundamental physics in two main arenas, namely the search for ultra-light dark matter and the search of GWs in the mid-frequency range between the peak sensitivity of the Earth- and space-based GW detectors relying on optical interferometry. The experimental development will deploy on Earth three generations of vertical AIs with increasing baselines, starting with 10 m (see scheme in Fig. \ref{fig:aion}-left), then 100 m, and finally 1 km. These successive phases will pave the way for a space mission, where a baseline of thousands of km is considered. As for the MAGIS experiment, AION will adopt for the atomic sensors strontium atoms coherently manipulated on the clock line at 698 nm, and several sensitivity enhanced techniques, as large momentum splitting, enhanced phase resolution, and resonant mode operation \cite{Graham2016}, to achieve the targeted performance.

Already the first generation of AION should have the sufficient sensitivity to test different ultra-light dark matter candidates at or slightly beyond the state-of-the-art; the second generation instrument will instead give significant contributions for the search of both dark matter and GWs from astrophysical sources and from the early universe. The networking capabilities of the instrument as GW detector have been studied taking in consideration its parallel operation with another instrument, as is the case of MAGIS: the result is a much improved pointing accuracy for the potential GW sources. Operating instead AION in synergy with detectors that probe other frequency bands will bring to multiband GW astronomy \cite{Sesana2016}, with complementary information for the same event coming from different instruments operated at different times in contiguous frequency regions, as shown in Fig. \ref{fig:aion}-right.

\subsection{VLBAI -- Very Long Baseline Atom Interferometry \label{ssec:vlbai}}

\begin{figure}[htb]
\centering
\includegraphics[width=6cm]{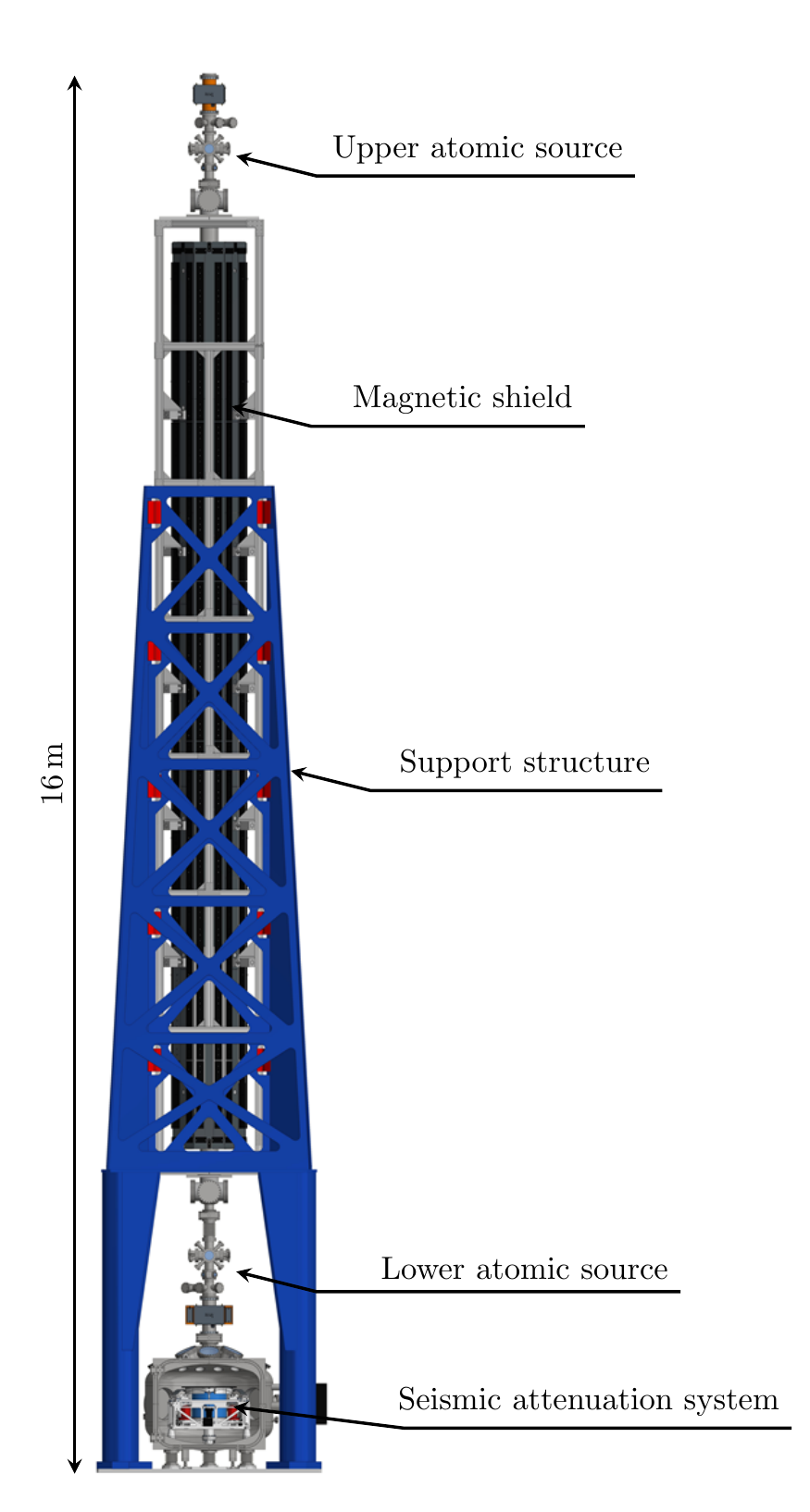}
\caption{Scheme of the VLBAI instrument, consisting in a 10 m tall atomic fountain for Yb and Rb atoms, being developed for precise tests of general relativity, quantum mechanics and their interplay. With permission of D. Schlippert.}
\label{fig:vlbai}
\end{figure}

The VLBAI experiment \cite{Schlippert2020} is implementing a 10 m tall atomic fountain in Hanover (Germany) to perform precise geodesy measurements, and several tests of fundamental physics, studying for example possible gravitationally-induced decoherence channels for quantum superposition states \cite{Pikovski2015} (Fig.\ref{fig:vlbai}). To reach these targets, the instrument will make use of state-of-the-art atom interferometry techniques, like LMT and delta-kick-cooling collimation \cite{Mntinga2013}, dual species atomic sources at both its sides, and enhanced control of the environmental disturbances. More specifically, vibrations will be mitigated through seismic attenuation systems based on geometric anti-springs \cite{Bertolini2000} and opto-mechanical devices \cite{Richardson2020}, whereas the spurious magnetic fields will be attenuated by a multi-layer $\mu$-metal shield. The expected performance in terms of gravity-gradient sensitivity will be 5$\times$10$^{-10}$ s$^{-2}$ at 1~s.

\section{Role of atom interferometry in GW astronomy}

The prospect of achieving strain sensitivities compatible with a detection in the infrasound band with atom interferometry responds to the need of widening the observation window opened by LIGO-Virgo in the gravitational spectrum. A future atom interferometry based GWD will be complimentary to the Earth based (LIGO-Virgo) and the planned space (LISA) optical interferometers, filling the sensitivity gap between them in the frequency spectrum. An open issue concerns the identification of potential GW sources in the specific bandwidth: astrophysical sources of GWs at $\approx 0.1$ Hz have been studied in \cite{Mandel2018,Kuns2020,Sedda2020}, and other more exotic sources have been taken into account like environmental effects from scalar fields and Dark Matter (DM) particles around merging compact objects \cite{Eda2015,Baumann2019}, cosmological phase transitions \cite{Weir2018}, and the formation of primordial black holes \cite{Saito2009}. Unlike for LISA where millions of BHBs and NS-NS inspirals will clog the observation band, astro-physical events will rather separately sweep the atom interferometry target band in the infrasound, opening towards the detection of stochastic GW background from the early evolution of the Universe \cite{Allen1999,Dimopoulos2008b,Lasky2016}.

\begin{figure}[t!]
  \centering
  \includegraphics[width=.7\textwidth]{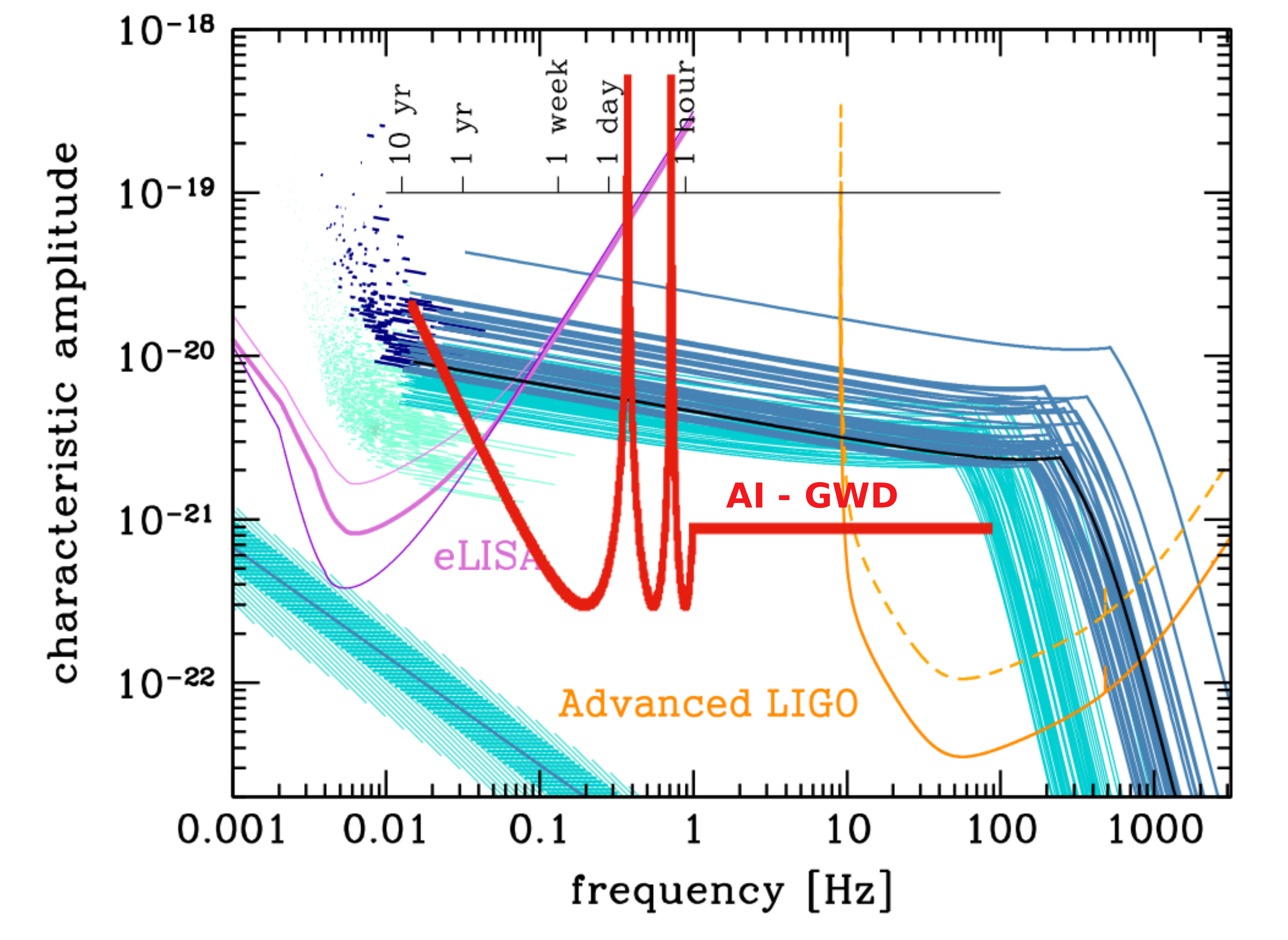}
  \caption{\label{fig:mbGWA} Multiband GW astronomy: the sensitivity curve of a future atom interferometry based GWD (red) is compared to the sensitivity curves of Adv-LIGO (yellow) and eLISA (violet). The blue curves represent characteristic amplitude tracks of BHB sources; the track completed by GW150914 is shown in black. Atom interferometry will fill the sensitivity gap between space and Earth based optical GWDs, opening the possibility to follow the inspiral phase of merging events that will be detected by the advanced-LIGO, or to see the merging of events at lower frequency. Reprinted and adapted figure with permission from \cite{Sesana2016}, \copyright $\,$ 2016 by the American Physical Society.}
\end{figure}

The potential role of atomic quantum sensors in GW astronomy is being increasingly studied. In Ref. \cite{Graham2018} the advantages given by a hypothetical atom interferometry based GWD -- on Earth or even better in space -- were considered for the specific GW150914 event, i.e. the first detection operated by LIGO-Virgo: the chirping GW frequency signal would have spent almost 10 months in the detection band, giving time and data for a much improved sky localization and a precise alert signal to the optical interferometers for the optimal detection of the merging. But other fascinating possibilities are being investigated in relation to both the possible GW sources and the so-called \textit{Multiband GW Astronomy}. The latter has been studied in \cite{Sesana2016} for the coupling of eLISA and Adv-LIGO: combining the two instruments will make possible precise gravity and cosmology tests by tracking compact binary inspirals from their early phase till the merging, and as well the mutual validation and calibration of the two instruments. Combining an atom interferometry based GWD with optical interferometry instruments (see Fig. \ref{fig:mbGWA}) will give multiple outcomes: (i) the coupling to the very low frequency eLISA will make the GWD relying on atomic quantum sensors the high frequency side instrument, targeting the merging of massive binary systems (10$^3$ to 10$^4$ M$_\odot$) which will never reach the LIGO-Virgo detection window; (ii) the coupling to LIGO-Virgo will instead give to the atom interferometry based GWD the task of tracking the inspiral phases of events later to be detected by the optical interferometer. Another possibility is to use an atom interferometry based GWD to mitigate the effect of GGN on Earth based optical GWD \cite{Chaibi2016}: adopting the two approaches on the same optical link would improve the sensitivity curve at low frequency for the optical instrument, specially if the AIs array geometry is adopted to map the GGN. It will be interesting to evaluate the impact of different generations of atom interferometry based GWD for this specific purpose, starting with the MIGA demonstrator and the first phase instruments planned in MAGIS, AION, and ZAIGA.

\subsection{GW sources in the atom interferometry detection bandwidth and multiband astronomy}

The target band for atom interferometry based GW detectors sits between the sensitivity curves of the operative optical detectors LIGO-Virgo and the foreseen LISA instrument \cite{Moore2014}, and will be then complimentary to them. Extensive studies of potential sources of GWs in the mid-band frequency have been already carried out \cite{Dimopoulos2008b,Mandel2018,Canuel2020,ElNeaj2020,Badurina2020}, and they indicate plenty of target signals of both astrophysical and cosmological origin.

The most abundant expected signals will come from compact binaries, composed by diverse systems like neutron stars (NSs), white dwarfs (WDs), and black holes (BHs). Depending on the specific binary system, and on its total mass, atom interferometry based GW detectors will be sensitive to different evolution phases: in the case of stellar-mass binaries, the inspiral phase will be measured, whereas for more massive systems like the so-called intermediate mass black holes (IMBH with masses between 10$^2 M_{\odot}$ and 10$^4 M_{\odot}$) the target phase will be the merger. The existence of the latter events has been recently corroborated by the detection of a 150$M_{\odot}$ BBH merger \cite{Abbott2020}, and for other reasons also by the discovery of a 70$M_{\odot}$ stellar-mass black hole \cite{Liu2019}. The availability of a decihertz observatory, as it was defined in \cite{Sedda2020} an instrument covering the frequency band around the 0.1 Hz, would be key in studying the evolution channels of the stellar-mass binary systems, shedding light on the formation mechanisms of the mergers events later detected by the LIGO-Virgo interferometers. The long observation time available to such detector would provide a wealth of complimentary information on the binary system, allowing a greatly improved parameter estimation; this is the case of the symmetric mass ratio between the two components of the binary system, which is one of the main indicators of its formation mechanism. Remarkably, the long observation would provide also an improved sky localization of the detected merger \cite{Mandel2018,Graham2018}, which can help in the determination of associated signals on non-gravitational channels, and for the definition of the local distance ladder via standard siren measurements \cite{Schutz1986}.

Other astrophysical events whose signatures are expected in the mid-frequency band are WD binary mergers, and notably type Ia supernovae, and the merging in the presence of a third body, which, as highlighted in \cite{Mandel2018}, would perturb the signal produced by a standard binary system in a measurable way at low frequency.

The detectors in the mid-frequency band would also boost the search for the signals of several cosmological sources, like first-order phase transitions in the early Universe \cite{Weir2018}, inflation, cosmic strings, and other effects resulting from extension of the Standard Model. Possible scientific discoveries awaiting in this context have been studied, within others, in \cite{ElNeaj2020,Graham2017,Badurina2020,Kuns2020}. Another stimulating opportunity will come from the network operation of two or more detectors in the band, as considered in \cite{Badurina2020}: by cross-correlating the output signals of far located instruments it will be possible to strongly mitigate the intrinsic technical noise of each detector, giving access to the underlying stochastic background of gravitational radiation \cite{Allen1999}, in a frequency band that looks particularly promising.

\section{Outlook}

\subsection{Roadmap to increase sensitivity}

Quantum sensors based on matter wave interferometry represent a new groundbreaking tool in fundamental and applied science, and offer novel roads and possibilities in several research contexts, ranging from tests of general relativity and quantum mechanics to metrology and inertial navigation, to mention a few. In GW detection, instruments based on ultracold atoms introduce several key innovations with respect to standard optical interferometers, which account for their potential advantages: (i) the use of nearly ideal quantum sensors in pure free fall is at the basis of the sensitivity curve extending at much lower frequency for terrestrial instruments; (ii) the re-configurable sensitivity function allows to quickly change the detector from broad band to narrowband, with no need to intervene on the instrument but just selecting a different sequence of manipulation pulses \cite{Kolkowitz2016,Graham2016}; (iii) the possibility to implement laser-noise-free detectors with only a single baseline \cite{Graham2013}; (iv) more than two sensors can be placed on the same optical link, which can be exploited to mitigate the GGN by measuring its spatial signature and statistically averaging it \cite{Chaibi2016}. Translating all these possibilities into a real instrument for the detection of GWs requires to push the state-of-the-art for all the available instrumental parameters by many orders of magnitude. Increasing the gravity-gradiometer baseline, a process already underway, linearly improves the sensitivity to strain. The atomic sources must be optimized as concerning their fluxes, thanks to faster and more efficient cooling and trapping techniques \cite{Stellmer2013,Urvoy2019,Chen2019,Naik2020}, to reduce the atom shot noise limit. The control of the atomic external degrees of freedom needs further improvements, to obtain smaller and better collimated sensors which maintain the interferometer contrast and reduce the impact of systematic effects - e.g. gravity-gradients, wavefront aberrations, Coriolis and mean field effects \cite{Hensel2021}. Atom optics that coherently manipulate the matter waves, and that ultimately allow the readout of the local phase on the optical link, need to be further scaled up, to obtain a larger sensitivity via the transfer of a larger number of photons \cite{Chiow2011,Gebbe2021}; the same target can be obtained adopting non-classical atomic states to run the interferometer, so as to beat the atom shot noise limit \cite{Lucke2011,Hosten2016,Cox2016,Salvi2018}. Novel protocols to improve the instrument bandwidth and shape the sensitivity function must be defined and perfected, as it is the case for the resonant mode interferometry \cite{Bishof2013,Graham2016}, the interleaved scheme \cite{Biedermann2013,Savoie2018} to achieve continuous measurements or even oversampling, and enhanced manipulation sequences exploiting quantum measurements \cite{Kohlhaas2015}. The transfer of techniques from optics - for example the exploitation of optical resonators to manipulate matter waves \cite{Hamilton2015,Riou2017,Xu2019} - must be pursued, and the same for the hybridization between atom and optical interferometers to obtain a mutual performance improvement via their combined operation. In addition to all that, the impact of systematic effects must be reduced to take advantage of the extremely high sensitivity intrinsically provided by quantum sensors, and this objective requires a many sided approach, which consists in studying the corrections introduced by high order terms describing the matter waves dynamics \cite{Bertoldi2019,Overstreet2021}, developing techniques to mitigate or cancel some backgrounds, as is the case of the gravity gradient \cite{Roura2017,DAmico2017,Overstreet2018,Bertoldi2019}, the Newtonian noise \cite{Chaibi2016}, and the wavefront aberrations \cite{Schkolnik2015}.
The last decade witnessed a manifold improvement in the performances of atom interferometry; the road to have sensors based on matter waves capable of detecting GWs is still long, but several research activities are pushing the limits. A new generation of detectors is redefining the concept of ``large instrument'' in matter wave interferometry, and several large baseline instruments are in constructions or being designed. The same instruments will be used to develop the specific techniques - e.g. data analysis, network operation, mitigation of backgrounds - that one day will be necessary to run the GW detectors based on matter waves, acting in this sense as demonstrator technologies. The full deployment of the intrinsic potential of atom interferometry will require one day a dedicated spatial mission, where satellites will provide the quietest possible environment, notably free of Newtonian noise; a key legacy is represented in this regard first by the terrestrial microgravity platforms for cold atoms \cite{Mntinga2013,Barrett2016,Condon2019,Becker2018,Lotz2020}, and second by several study missions \cite{Aguilera2014,Trimeche2019,Battelier2019} and ongoing projects exploiting quantum atomic sensors in space \cite{Laurent2015,Aveline2020}.

\subsection{ELGAR -- European Laboratory for Gravitation and Atom-interferometric Research \label{ssec:elgar}}

We now finish this chapter by detailing a ground based GW detector project based on some the advanced techniques listed in the previous section: ELGAR~\cite{Canuel2020} plans to build an underground infrastructure based on atom interferometry, to study space-time and gravitation with the primary goal of detecting GWs in the infrasound band, from  0.1~Hz to 10~Hz. The instrument aims to become the cornerstone of multiband GW astronomy by covering a frequency band which is complementary to the third generation detector ET~\cite{Punturo2010}: if operated in parallel, the two detectors could provide the best sensitivity from 0.1 Hz  to 100 Hz. The further addition of the space antenna LISA~\cite{Jennrich2009} would enlarge the band with almost no blind spot from 1~mHz to 100~Hz. The ELGAR initiative is sustained since 2014 by an active research consortium that gathers 60 scientists from more than 20 laboratories and research groups in 6 European countries. Different sites are under consideration for the installation of the detector including Sos-Enattos in Sardinia~\cite{Naticchioni_2020} which is also a candidate site for the Einstein Telescope, and the LSBB laboratory~\cite{gaffet2019}, where MIGA is located, which is one of the best underground infrastructures in Europe in terms of ambient noise. 

\begin{figure}[htp]
\centering
\includegraphics[width=1\linewidth]{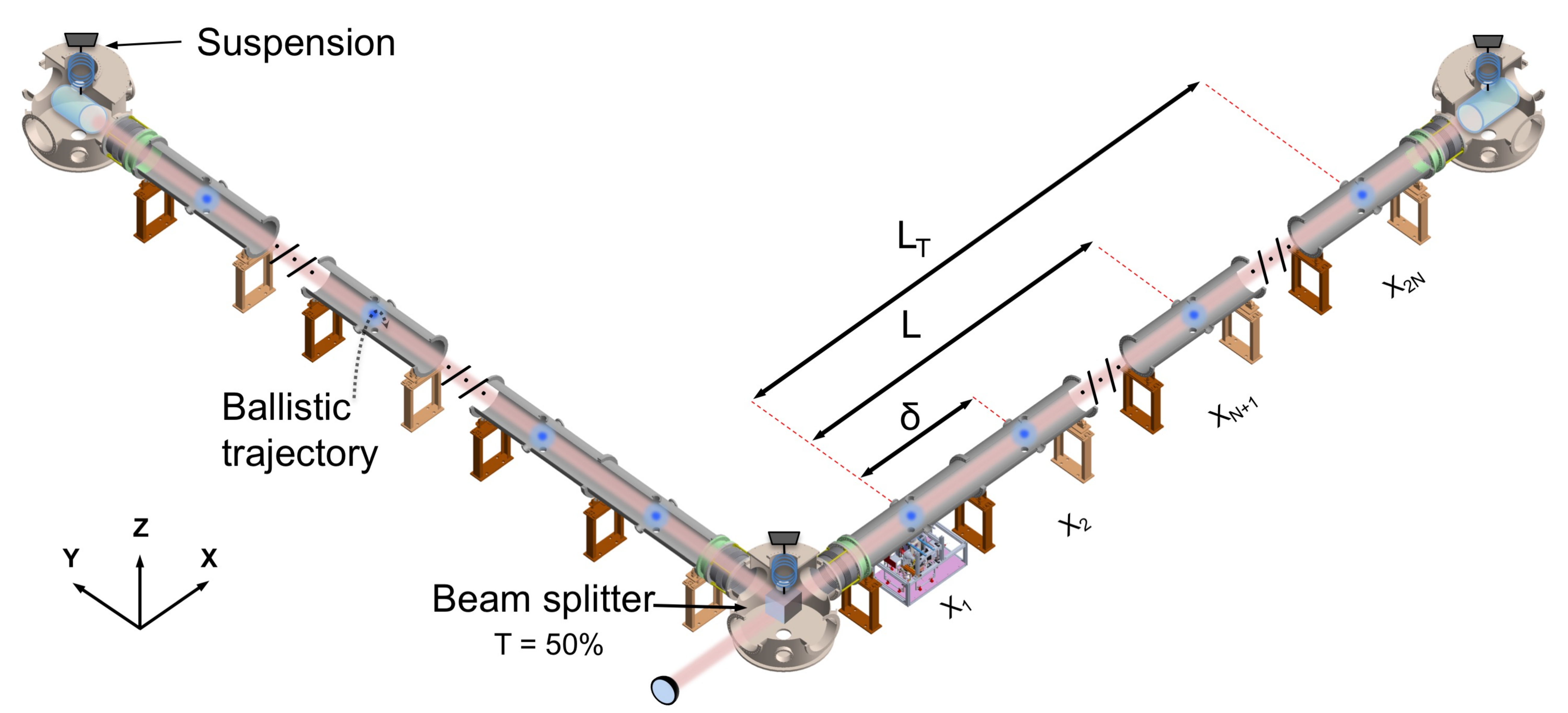}
\caption{Preliminary design of the ELGAR infrastructure, formed by a 2D array of $N$ gradiometers of baseline $L$, spaced regularly by a distance $\delta$ over a total length $L_T$. Taken from Ref.~\cite{Canuel2020} under CC BY 4.0. \label{fig:ElgarGeo}}
\end{figure}

The preliminary design of the antenna was published in Ref.~\cite{Canuel2020, ELGARtecno2020} and can be seen in Fig.~\ref{fig:ElgarGeo}.
ELGAR uses a detector geometry presented in Sec. \ref{sec:GGN}: a common laser interrogates two perpendicular arms of total length $L_T=$ 32.1~km, and each arm is composed of $N=80$ gradiometers of length $L=$ 16.3~km, regularly spaced by a distance $\delta=$ 200~m.

\begin{figure}[htp]
    \centering
    \includegraphics[width=1\linewidth]{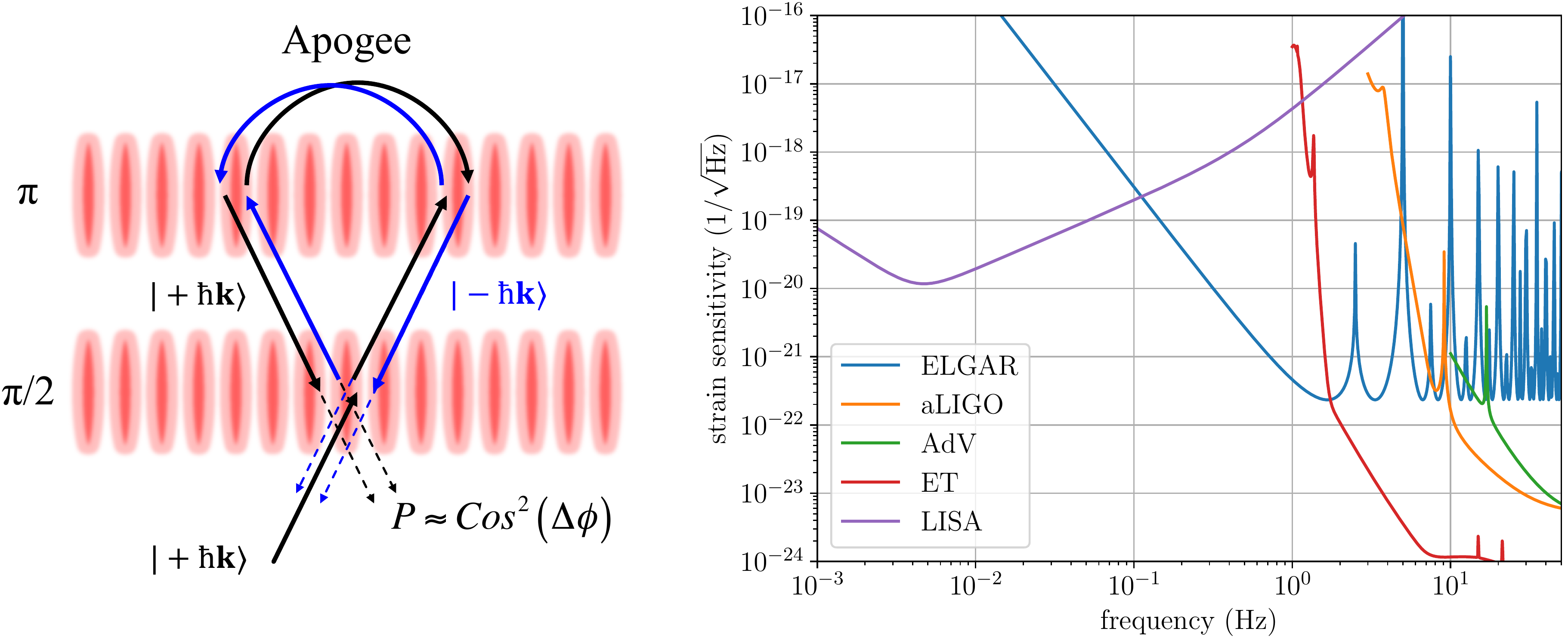}
    \caption{(left) Geometry of the ELGAR four pulse AI. The atoms experience a first $\pi/2$ pulse creating a superposition between two momentum states. Two successive $\pi$ pulses reverse the momentum states and exchange the matter-wave trajectories twice. Finally, a second $\pi/2$ pulse closes the interferometer. (right) Strain sensitivities for different gravitational wave detectors, which include ELGAR (atom shot noise limit), aLIGO, ET and LISA, and cover the frequency range from 10 mHz to 10 Hz. The resonances of the ELGAR sensitivity curve are linked to the windowing effect of the interrogation process and can effectively removed by changing the pulse sequence to a broadband mode \cite{ELGARtecno2020}. Taken from \cite{Canuel2020} under CC BY 4.0. \label{schematicexp2BOOK}}
\end{figure}

The antenna foresees to use different laser cooling techniques for producing $^{87}$Rb atom ensembles at a kinetic temperature of $\approx100$~pK, with a cloud size of about 16~mm, large enough to mitigate atom-atom interactions. After the cooling stage, the atoms are launched on a vertical parabolic trajectory towards the interferometric zone shown in Fig.~\ref{schematicexp2BOOK} (left) where they are coherently manipulated using a 4 pulses ``$\pi/2$-$\pi$-$\pi$-$\pi/2$" sequence~\cite{Audretsch1994}. This ``butterfly" configuration is chosen for its performance against spurious phase variations.
The coherent manipulation protocol for the atoms includes LMT~\cite{Chiow2011,Gebbe2021} by using a combination of Bragg diffraction~\cite{Audretsch1994} and Bloch oscillation~\cite{Dahan1996}, to reach $2n=$~1000 photon exchange at each pulse. Considering a flux of $10^{\text{12}}$~atoms/s and an integration time of $4T=800$~ms, the atom shot noise limited strain sensitivity of a single gradiometer antenna becomes $4.1 \times 10^{-21}/\sqrt{\rm{Hz}}$ at 1.7 Hz and improves to about $3.3 \times 10^{-22}/\sqrt{\rm{Hz}}$ at 1.7~Hz considering the response of the whole detector given by the average signal of Eq.~(\ref{average}). The strain sensitivity curve of the antenna can be seen in Fig.~\ref{schematicexp2BOOK} (right) together with other actual and future GW detector such as aLIGO, AdV, ET and LISA. We observe that ELGAR will cover the frequency gap between 100 mHz and 1 Hz between the sensitivity of optical ground and space based detectors.

\bibliographystyle{unsrtnat}

\end{document}